\documentclass[aps,prc,floatfix,showpacs,twocolumn]{revtex4-1}
\usepackage{ulem}
\usepackage{dcolumn}
\usepackage{bm}
\usepackage{color}
\usepackage{amssymb}
\usepackage{amsmath}
\usepackage{graphicx}
\usepackage{amsfonts}
\usepackage{slashed}
\usepackage{pstricks}
\usepackage{float}
\usepackage{hyperref}
\usepackage{array}
\allowdisplaybreaks
\usepackage{dcolumn}
\usepackage{epsf}
\usepackage{slashed}
\def\ra{\rangle}
\def\la{\langle}

\begin{document}
\title{Neutrino-nucleus cross section within the extended factorization scheme}
\author{
{Noemi} Rocco$^{\, {\rm a,b,c} }$,
{Carlo} Barbieri$^{\, {\rm a} }$,
{Omar} Benhar$^{\, {\rm d,e} }$,
{Arturo} De Pace$^{\, {\rm f} }$,
{Alessandro} Lovato$^{\, {\rm b,g} }$
}
\affiliation{
$^{\,{\rm a}}$\mbox{Department of Physics, University of Surrey, Guildford, GU2 7HX, UK}\\
$^{\,{\rm b}}$\mbox{Physics Division, Argonne National Laboratory, Argonne, Illinois 60439, USA}\\
$^{\,{\rm c}}$\mbox{Theoretical Physics Department, Fermi National Accelerator Laboratory, P.O. Box 500, Batavia, IL 60510, USA}\\
$^{\,{\rm d}}$\mbox{{Dipartimento di Fisica,``Sapienza'' Universit\`a di Roma, I-00185 Roma, Italy}}\\
$^{\,{\rm e}}$\mbox{{INFN, Sezione di Roma, I-00185 Roma, Italy}}\\
$^{\,{\rm f}}$\mbox{{INFN, Sezione di Torino, Via P. Giuria 1, I-10125 Torino, Italy}}\\
$^{\,{\rm g}}$\mbox{INFN-TIFPA Trento Institute of Fundamental Physics and Applications, Via Sommarive, 14, 38123 {Trento}, Italy}\\
}
\date{\today}

%
%%%%%%%%%%%%%%%%%%%%%%%%%%%%%%%%%%%%%%%%%%%%%%%%%%%%%%%%%%
\date{\today}
%%%%%%%%%%%%%%%%%%%%%%%%%%%%%%%%%%%%%%%%%%%%%%%%%%%%%%%%%%%%%%%%%%%%%%%%%
\begin{abstract} 
The factorization scheme, based on the impulse approximation and the spectral function formalism, has been recently generalized
to allow the description of electromagnetic nuclear interactions driven by two-nucleon currents. We have extended this framework to the case
of weak charged and neutral currents, and carried out calculations of the double-differential neutrino-carbon and neutrino-oxygen  cross 
sections using two different models of the target spectral functions. The results, showing a moderate dependence on the input spectral function, 
confirm that  our approach provides a consistent treatment of all reaction mechanisms contributing to the signals detected  by accelerator-based 
neutrino experiments.
\end{abstract}
\maketitle
%%%%%%%%%%%%%%%%%%%%%%%%%%%%%%%%%%%%%%%%%%%%%%%%%%%%%%%%%%%%%%%%%%%%%%%%%
\section{Introduction}
Accurate predictions of neutrino-nucleus interactions are pivotal to the success of the long-baseline neutrino-oscillation program. 
Current-generation~\cite{mb_web,nova_web,t2k_web,minerva_web} and next-generation~\cite{dune_web,hk_web}
experiments are sensitive to a broad range of energy, in which different reaction mechanisms, involving both nucleon and
nuclear excitations, are at play~\cite{Katori:2016yel,Benhar:2015wva}. At energies of the order of hundreds of MeVs, the leading mechanism is
quasielastic scattering, in which the probe interacts primarily with individual nucleons bound inside the nucleus. Corrections to this leading mechanism 
arise from processes in which the  lepton couples to interacting nucleons, either via nuclear correlations or two-body currents. Neutrinos can also 
excite a struck nucleon to a baryon resonance state that quickly decays into pions, or give rise to deep-inelastic scattering (DIS) processes.

Constructing a framework suitable to consistently describe neutrino-nucleus interactions in the broad energy regime relevant for neutrino-oscillation
experiments is a formidable nuclear-theory challenge. Nuclear EFTs, which provide a way to systematically construct nuclear interactions 
and currents within the framework of a low-momentum expansion, can be safely applied to describe ground-state properties of the target nucleus. 
On the other hand, because of the large energy- and momentum-transfer involved, their extent of applicability to model the interaction vertex and 
the final hadronic states, where relativistic effects cannot be neglected, is more questionable. Hence, it is of paramount 
importance to validate theoretical predictions for neutrino-nucleus scattering through a systematic comparison with the large body of available 
electron scattering data~\cite{Benhar:2011ef}. In fact, the ability to explain electron scattering experiments should be seen as an obvious
prerequisite, to be met by any models of the nuclear response to weak interactions~\cite{Benhar:2016jkq}.

Green's function Monte Carlo (GFMC)~\cite{Carlson:2014vla} and Self-Consistent Green's function (SCGF) approaches~\cite{Dickhoff2004ppnp,Barbieri2014QMBT} 
are suitable to perform accurate calculations of atomic nuclei, starting from the individual interactions among their constituents. 
Up to moderate values of the momentum transfer, the electroweak responses obtained
within GFMC in the quasielastic sector are virtually exact and give full account of initial and final state correlations, and electroweak 
two-body currents~\cite{Lovato:2016gkq,Lovato:2017cux}. Once relativistic effects in nuclear kinematics are included, an excellent agreement with electron scattering data off $^4$He has 
been found~\cite{Rocco:2018tes}. Because of the exponential scaling with the number of nucleons, it is unlikely that GFMC will be applied to compute
the electroweak responses of nuclei larger than $^{12}$C in the near future. In addition, the use of integral-transform techniques precludes a proper treatment of the 
energy dependence of the current operators, particularly important at energies higher than those corresponding to the quasielastic kinematics. 
Finally, despite encouraging preliminary results have recently been obtained~\cite{Madeira:2018ykd}, the explicit inclusion of pions -- and hence a proper description of the 
resonance region -- are still a long way ahead. 
The SCGF belongs to a class of polynomially-scaling many-body methods that allow to reach nuclei with mass number $A$ up to $\sim$100  
with relatively modest computational costs. Within this framework, the two-body (particle-hole) polarization propagator provides information on the transition to 
low-energy excited final states, relevant for the giant-resonance region~\cite{Brand1988ERPA,Barbieri2003o16ex,Raimondi2017GR}. The one-body propagator, instead, is directly linked to the hole spectral function (SF) that gives a detailed account on the energy- and momentum distribution of bound nucleons inside the target~\cite{Dickhoff2004ppnp}.

The formalism based on the impulse approximation (IA) and realistic hole SFs allows to combine a realistic description of the initial
state of the nuclear target with a fully-relativistic interaction vertex and kinematics~\cite{Benhar:2006wy}. Calculations carried out employing hole SF computed within the
correlated-basis function (CBF) and the SCGF theories have been extensively validated against electron-nucleus scattering data on a number of nuclei~\cite{Ankowski:2014yfa,Benhar:2005dj,Rocco:2018vbf,Middleton2010pn}. The 
somewhat  oversimplified treatment of final-state interactions (FSI) to which the struck nucleon undergoes has been corroborated comparing the 
electromagnetic response functions of $^{12}$C from CBF with those of the GFMC~\cite{Rocco:2016ejr}.

More recently, the factorization scheme underlying IA and the SF formalism has been generalized to include electromagnetic relativistic meson-exchange two-body currents (MEC), arising 
from pairs of interacting nucleons~\cite{Benhar:2015ula}. Employing nuclear overlaps and consistent SFs obtained within the CBF theory, the authors of 
Refs.~\cite{Rocco:2015cil} have analyzed the role of MEC in electron scattering off $^{12}$C. They found that two-body currents are mostly effective in 
the ``dip'' region, between the quasielastic and the $\Delta$-production peaks. Their inclusion appreciably improves the agreement between 
theory and data. 

In this work, we further extend the IA scheme by introducing the MEC relevant for charged-current (CC) and neutral-current (NC) interactions. We study 
their role in neutrino and anti-neutrino scattering off $^{12}$C and $^{16}$O nuclei, both used as targets in neutrino-oscillation experiments. We adopt the 
two-body currents derived in Ref.~\cite{Simo:2016ikv} from the weak pion-production model of Ref.~\cite{Hernandez:2007qq}. It has been shown that they provide results consistent with 
those of Ref.~\cite{DePace:2003}, which were also adopted in the extension of the IA and SF formalism of Ref.~\cite{Rocco:2015cil}.

We develop a dedicated code that automatically carries out the calculation of the MEC spin-isospin matrix elements, performing the integration using the
Metropolis Monte Carlo algorithm~\cite{Metropolis:1953am}. To validate our implementation of the two-body currents, we perform a benchmark calculation of the CC response 
functions within the relativistic Fermi gas model, comparing our results with the findings of  Ref.~\cite{Simo:2016ikv}. 

We consider two nuclear SFs, derived within the framework of nuclear many-body theory using the CBF formalism~\cite{Benhar:1994} and the self-consistent Green's function theory~\cite{Dickhoff2004ppnp,Barbieri2017LNP}.
These two approaches start from different, albeit realistic, nuclear hamiltonians to describe the interactions between protons and neutrons.
Moreover, the approximations  involved in the calculations of the hole spectral function are also peculiar to of each of the two methods. Hence, a comparison of the
cross sections obtained employing the CBF and the SCGF nuclear SFs helps gauging the theoretical error of the calculation. 

More specifically, we analyze the double-differential cross sections of $^{12}$C and $^{16}$O for both CC and NC transitions for incoming (anti)neutrino energy
of $1$ GeV and two values of the scattering angle: $\theta_\mu=30^\circ$ and $\theta_\mu=70^\circ$. We also present results for the total CC cross 
section for neutrino and anti-neutrino scattering off $^{12}$C as a function of the incoming (anti)neutrino energy. Our calculations are compared with the experimental 
data extracted by the MiniBooNE collaboration~\cite{AguilarArevalo:2010zc}.

The structure of the nuclear cross section, as well as its expression in terms of relevant response functions are reviewed in Section~\ref{sec:formalism}. 
Section~\ref{sec:impulse} is devoted to the description of the IA, including its extension to account for a consistent treatment of one- and two-nucleon
current contributions. The CBF theory and SCGF approaches are also briefly outlined. In Section~\ref{sec:currents} we discuss the explicit expressions 
of the relativistic two-body currents employed, while Section~\ref{sec:numerical} is dedicated to their numerical implementation. In Section~\ref{sec:results}
we present our results and in Section~\ref{sec:conclusions} we state our conclusions. 

\section{Formalism}
\label{sec:formalism}
The double-differential cross section for $\nu$ and $\bar{\nu}$ inclusive scattering off a nucleus can be expressed as~\cite{Shen:2012,rocco:notes}
\begin{align}
\Big(\frac{d\sigma}{dT^\prime d\cos\theta^\prime}\Big)_{\nu/\bar{\nu}}&=\frac{G^2}{2\pi}\frac{k^\prime}{2 E_\nu}\Big[\hat{L}_{CC}R_{CC}+2\hat{L}_{CL}R_{CL}\nonumber\\
&+\hat{L}_{LL}R_{LL}+\hat{L}_{T}R_{T}\pm 2 \hat{L}_{T^\prime}R_{T^\prime}\Big]\ ,
\label{eq:cross_sec}
\end{align}
where $G=G_F$ and $G=G_F \cos\theta_c$ for NC and CC processes, respectively, with $\cos\theta_c=0.97425$~\cite{PDG}. The $+$ $(-)$ sign corresponds to $\nu$ ($\bar{\nu}$)
induced reactions. We adopt the value $G_F = 1.1803 \times 10^{-5}\,\rm GeV^{-2}$, as from the analysis of $0^+ \to 0^+$ nuclear $\beta$-decays of Ref.~\cite{Herczeg:1999}, which 
accounts for the bulk of the inner radiative corrections~\cite{Nakamura:2002jg}.
With $k=(E_\nu,\mathbf{k})$ and $k^\prime=(E_\ell,\mathbf{k}^\prime)$ we denote the initial neutrino and the final lepton four-momenta, respectively, and $\theta$ 
is the lepton scattering angle. Introducing the four-momentum 
\begin{equation}
Q=k+k^\prime=(\Omega, {\bf Q}) \, , \quad {\bf Q}= (Q_x,0,Q_z)
\end{equation}
and the momentum transfer
\begin{equation}
q=k-k^\prime=(\omega, {\bf q})\, , \quad {\bf q}= (0,0,q_z),
\end{equation}
the kinematical factors can be conveniently cast in the form
\begin{align}
\hat{L}_{CC}&={\Omega}^2-q_z^2 -m_\ell^2\nonumber\\
\hat{L}_{CL}&=(-\Omega Q_z+\omega q_z)\nonumber\\
\hat{L}_{LL}&= {Q_z}^2 - \omega^2 + m_\ell^2\nonumber\\
\hat{L}_{T}&= \frac{{Q_x}^2}{2}- q^2 + m_\ell^2 \nonumber\\
\hat{L}_{T^\prime}&= \Omega q_z- \omega Q_z\, ,
\end{align}
with $m_\ell^2=k^{\prime\, 2}$ being the mass of the outgoing lepton. 
The five electroweak response functions are given by
\begin{align}
R_{CC}&=W^{00}\nonumber\\
R_{CL}&=-\frac{1}{2}(W^{03}+W^{30})\nonumber\\
R_{LL}&=W^{33}\nonumber\\
R_T&=W^{11}+W^{22}\nonumber\\
R_{T^\prime}&=-\frac{i}{2}(W^{12}-W^{21})\,,
\end{align}
where the hadronic tensor
\begin{align}
W^{\mu\nu}=\sum_f \langle 0|j^{\mu \, \dagger}|f\rangle \langle f| j^\nu |0 \rangle \delta (E_0+\omega -E_f)
\label{eq:had_tens}
\end{align}
contains all information on the structure of the target. It is defined in terms of the transition between the initial and final nuclear 
states $|0\rangle$ and $|f\rangle$, with energies $E_0$ and $E_f$, induced by the nuclear current operator $j^\mu$.

Note that the sum in Eq.\eqref{eq:had_tens} includes the contributions of inelastic processes, leading to 
the appearance of hadrons other than nucleons in final state, which we will not discuss in this article. The derivation of  
the inelastic neutrino-nucleus cross section within the SF formalism can be found in Ref.~\cite{Vagnoni:2017hll}.

\section{Impulse Approximation}
\label{sec:impulse}
At relatively large values of the momentum transfer, typically $|{\bf q}| \gtrsim 500$ MeV, the impulse approximation (IA) can be safely applied
under the assumption that the struck nucleon is decoupled from the spectator (A-1) particles~\cite{Benhar:2006wy, Benhar:2015wva}.
Within the IA, the nuclear current operator reduces to a sum of
one-body terms, $j^\mu=\sum_i j^\mu_i$ and the nuclear final state factorizes as
\begin{align}  
|\psi_f^A \rangle \rightarrow |p\rangle \otimes |\psi_f^{A-1}\rangle\, .
\end{align}
In the above equation $|p\rangle$ denotes the final-state nucleon with momentum ${\bf p}$ and energy $e(\bf p)$, while $|\psi_f^{A-1}\rangle$
describes the \hbox{$(A-1)$-body} spectator system. Its energy and recoiling momentum are fixed by energy and momentum conservation
\begin{align}
E_f^{A-1}=\omega +E_0-e({\bf p})\, ,\quad {\bf P}^{A-1}_f={\bf q}-{\bf p}\, .
\end{align}

Employing the factorization ansatz and inserting a single-nucleon completeness relation, the matrix element of the current operator can be written as
\begin{align}
&\langle \psi_f^A|j^{\mu} |\psi_0^A \rangle \rightarrow \sum_k [\langle \psi_f^{A-1}| \otimes \langle k | ] \, | \psi_0^A\rangle  \langle p | \sum_i j^\mu_i | k \rangle\,.
\label{fact:1b}
\end{align}
Substituting the last equation in Eq.~\eqref{eq:had_tens}, the incoherent contribution to the hadron tensor, dominant at large 
momentum transfer, is given by
\begin{align}
&W^{\mu\nu}_{\rm 1b}({\bf q},\omega)=\nonumber\\
&\quad \sum_{p,k,f}\sum_{i}\, \langle k | {j_{i}^\mu}^\dagger |p \rangle \langle p |  j_{i}^\nu | k \rangle
| \langle \psi_0^A| [ | \psi_f^{A-1}\rangle \otimes |k \rangle ]  |^2\nonumber\\
&\quad \times   \delta(\omega-e(\mathbf{p}) - E_{f}^{A-1} +E^A_0)\, ,
\end{align}
where the subscript ``1b'' indicates that only one-body currents have been included. 
Using the identity
\begin{align}
\delta(\omega&-e({\bf p})-E^{A-1}_f+E^A_0)=\nonumber\\
&   \int dE\, \delta(\omega+E-e({\bf p})) \, \delta(E+E^{A-1}_f-E^A_0)  \; ,
\end{align}
and the fact that momentum conservation in the single-nucleon vertex implies ${\bf p}={\bf k}+{\bf q}$,
we can rewrite the hadron tensor as
\begin{align}
W^{\mu\nu}_{\rm 1b}({\bf q},\omega)=& \int \frac{d^3k}{(2\pi)^3} dE P_h({\bf k},E)\frac{m_N^2}{e({\bf k})e({\bf k+q})}\nonumber\\
&\times  \sum_{i}\, \langle k | {j_{i}^\mu}^\dagger |k+q \rangle \langle k+q |  j_{i}^\nu | k \rangle\nonumber\\
& \times \delta(\omega+E-e(\mathbf{k+q}))\, .
\label{had:tens}
\end{align}
The factors  $m_N/e({\bf k})$ and $m_N/e({\bf k+q})$, $m_N$ being the mass of the nucleon, are included to 
account for the implicit covariant  normalization of the four-spinors of the initial and final nucleons in the matrix elements 
of the relativistic current.

The hole spectral function
\begin{align}
P_h(\mathbf{k},E)&=\sum_f |\langle \psi_0^A| [|k\rangle \otimes |\psi_f^{A-1}\rangle]|^2\nonumber\\
&\times\delta(E+E_{f}^{A-1}-E^A_0)
\label{pke:hole}
\end{align}
provides the probability distribution of removing a nucleon with momentum ${\bf k}$ from the target nucleus, 
leaving the residual $(A-1)$-nucleon system with an excitation energy $E$.
Note that in Eq.~\eqref{had:tens} we neglected Coulomb interactions and the other (small) isospin-breaking
terms and made the assumption, largely justified in the case of symmetric closed shell nuclei, 
that the proton and neutron spectral functions are identical.

Using the Sokhotski-Plemelj theorem~\cite{weinberg1995quantum} we can rewrite Eq.~\eqref{pke:hole} as
\begin{align}
P_h(\mathbf{k},E)&=\frac{1}{\pi}\sum_f\text{Im}\langle \psi_0^A| \frac{1}{E+E_{f}^{A-1}-E_0^A-i\epsilon} [|k\rangle\nonumber\\
&\otimes |\psi_f^{A-1}\rangle][ \langle\psi_f^{A-1}|\otimes\langle k|] |\psi_0^A\rangle\, .
\end{align}
Exploiting the fact that $H|\psi_f^{A-1}\rangle=E_{f}^{A-1}|\psi_f^{A-1}\rangle$ and the completeness of the $A-1$ states, the
hole SF can be expressed in terms of the hole Green's function 
\begin{align}
P_h(\mathbf{k},E)=\frac{1}{\pi}\text{Im}\langle \psi_0^A| a_{\mathbf{k}}^\dagger  \frac{1}{E+(H-E_0^A)-i\epsilon} a_{\mathbf{k}}|\psi_0^A\rangle\, .
\label{pke_vs_gf}
\end{align}
Finally, it has to be noted that the single nucleon momentum distribution, corresponds to the integral of the spectral function over
the removal energy
\begin{align}
n({\bf k})=\langle\psi_0^A|a^\dagger_k a_k|\psi_0^A\rangle= \int dE P({\bf k},E)\, .
\end{align}

In the kinematical region in which the interactions between the struck particle and the spectator system cannot be neglected,
the IA results are modified to include the effect of final-state interactions (FSI). The multiple scatterings that the struck particle 
undergoes during its propagation through the nuclear medium can be taken into account through a convolution 
scheme~\cite{Benhar:1991af,Ankowski:2014yfa}, which amounts to integrating the IA prediction with a folding function that describes
the effects of FSI between the struck particle and the $A-1$ spectator system. In addition, to describe the propagation of the knocked-out
particle in the mean-field generated by the spectator system, the energy spectrum of the knocked-out nucleon is modified with the real
part of an optical potential derived from the Dirac phenomenological fit of Ref~\cite{Cooper:1993nx}.

In this work, aimed at devising the formalism for including relativistic meson-exchange currents within two realistic models
of the nuclear ground-state, FSI are disregarded. On the other hand, we will fully account them in the forthcoming calculations of the 
flux-integrated double-differential neutrino-nucleus cross sections.

\subsection{Correlated basis function theory}
Consistently with the spectral representation of the two-point Green's function, the CBF hole SFs of $^{12}$C and $^{16}$O  are 
written as the sum of two contributions~\cite{Benhar:1994hw}:
\begin{align}
P_h({\bf k},E)= P^{ 1h}_h({\bf k},E) + P_h^{\rm corr}({\bf k},E)  \label{eq:fullPh} \ .
\end{align}

The one-hole term is obtained from a modified mean-field scheme
\begin{align}
\label{Pke:MF}
P^{1h}_h({\bf k},E) = \sum_{\alpha\in \{{\rm F}\}} Z_\alpha |\phi_\alpha({\bf k})|^2 F_\alpha(E-e_\alpha) \ , 
\end{align}
where the sum runs over all occupied single-particle nuclear states, labeled by the index $\alpha$, and $\phi_\alpha({\bf k})$ 
is the Fourier transform of the shell-model orbital with energy $e_\alpha$. The {\it spectroscopic}
factor $Z_\alpha < 1$ and the function $F_\alpha(E-e_\alpha)$, describing the energy width of the state $\alpha$,
account for the effects of residual interactions that are not included in the mean-field picture. In the absence of
residual interactions, $Z_\alpha \to 1$  and $F_\alpha(E-e_\alpha) \to \delta_\alpha(E-e_\alpha)$. The spectroscopic
factors and the widths of the $s$ and $p$ states of $^{12}$C and $^{16}$O have been taken from the analysis
of $(e,e^\prime p)$ data carried out in Refs.~\cite{Mougey:1976sc,Turck:1981,Dutta:2003yt}.

The correlated part of the SF for finite nuclei $P_h^{\rm corr}({\bf k},E)$ is obtained through local density approximation (LDA) 
procedure
\begin{align}
\label{Pke:corr}
P^{ \rm corr}_h({\bf k},E) = \int d^3R \  \rho_A({\bf R}) P^{\rm corr}_{h,\,NM}({\bf k},E; \rho_A({\bf R}))\,,
\end{align}
In the above equation, $\rho_A(\mathbf{R})$ is the nuclear density distribution of the nucleus and $P^{\rm corr}_{h\,,NM}({\bf k},E; \rho)$ is
the correlation component of the SF of isospin-symmetric nuclear matter at density $\rho$. The use of the LDA to account for 
$P^{\rm corr}_h({\bf k},E)$ is based on the premise that short-range nuclear dynamics are largely unaffected by surface and shell 
effects. 

CBF calculations of the hole SF in isospin-symmetric nuclear matter are carried out considering overlaps involving 
the ground-state and one-hole and two-holes-one-particle excitations in $|\psi_f^{A-1}\rangle$~\cite{Benhar:1989aw, Benhar:1994hw}. 
They are consistently obtained from the following set of correlated basis (CB) states
\begin{equation}
\label{corr:st}
|\psi_n\ra_{\rm CB}= \frac{{\mathcal F} |\Phi_n\ra}{\la \Phi_n |{\mathcal F}^\dagger {\mathcal F}|\Phi_n \ra^{1/2}} \ , 
\end{equation}
where $|\Phi_n\ra$ is an independent-particle state, generic eigenstate of the free Fermi gas Hamiltonian, and the many-body 
correlation operator ${\mathcal F}$ is given by
\begin{equation}
{\mathcal F}= {\mathcal S} \Big[ \prod_{j>i=1}^A F_{ij} \Big] \ .
\end{equation}
The form of the two-body correlation operator $F_{ij}$ reflects the complexity of realitistic NN potential~\cite{Wiringa:1994wb}
\begin{equation}
F_{ij}= \sum_{n=1}^6 f^n(r_{ij})O^n_{ij}\ ,
\end{equation}
with $r_{ij}= |{\bf r}_i - {\bf r}_j|$ and
\begin{equation}
\label{oper}
O^{n\leq 6}_{ij}= [1, ({\bm \sigma}_i\cdot {\bm \sigma}_j),S_{ij} ]\otimes [1, ({\bm \tau_i}\cdot{\bm \tau_j})] \ ,
\end{equation}
In the above equation, ${\bm \sigma}_i$ and ${\bm \tau}_i$ are Pauli matrices acting in the spin and isospin space, respectively, and $S_{ij}$ is the tensor operator given by
\begin{equation}
S_{ij}= \frac{3}{r^2_{ij}}({\bm \sigma}_i\cdot {\bf r}_{ij})({\bm \sigma}_j \cdot{\bf r}_{ij})- ({\bm \sigma}_i\cdot {\bm \sigma}_j)\ .
\end{equation}
The CB states are first orthogonalized (OCB)~\cite{Fantoni:1988zz} preserving, in the thermodynamical limit, the diagonal matrix 
elements between CB states. Then, standard perturbation theory is used to express the eigenstates of the nuclear Hamiltonian
in terms of the OCB. Any eigenstate has a large overlap with the $n-$hole-$m-$particle OCB and hence perturbation theory in
this basis is rapidly converging. 

The nuclear-matter SF can be conveniently split into two components, displaying distinctly different energy 
dependences~\cite{Benhar:2006wy, Benhar:2015wva, Benhar:1990zz, Benhar:1994hw}. The single-particle one, associated
to one-hole states in $|\psi_f^{A-1}\rangle$ of Eq.~(\ref{pke:hole}), exhibits a collection of peaks corresponding to the energies
of the single-particle states belonging to the Fermi sea. The continuum, or correlation, component corresponds to states involving
at least two-hole\textendash one-particle contributions in $|\psi_f^{A-1}\rangle$. Its behavior as a function of $E$ is smooth
and it extends to large values of removal energy and momentum~\cite{Benhar:1989aw}. It has to be noted that the correlated
part would be strictly zero if nuclear correlations were not accounted for. As a consequence, the energy-dependence exhibited by 
$P^{\rm corr}_h({\bf k},E)$, showing a widespread background extending up to large values
of both $k$ and  $E$, is completely different from that of $P^{ 1h}_h({\bf k},E)$. For $k>p_F$, $P^{\rm corr}_h({\bf k},E)$
coincides with $P_h({\bf k},E)$ and its integral over the energy gives the so-called continuous part of the momentum distribution.

\subsection{Self-consistent Green's function}
\label{sec:SCGF}
The SCGF approaqch is appealing to our purposes because the hole component of the one-body
Green's function, which is the central quantity of the formalism, is directly related to $P_h({\bf k},E)$ through
Eq.~\eqref{pke_vs_gf}.  This allows to compute the complete spectral function directly form {\it ab initio}
theory and without a priori assumptions on the form of correlations.

The one-body Green's function is written as the sum of a forward-going ($g_{\alpha\beta}^>(\omega)$) and 
a backward-going ($g_{\alpha\beta}^>(\omega)$) terms that describe the propagation  of a particle and a hole state, respectively~\cite{Dickhoff2008book}. 
In the so-called Lehmann representation, this reads:
\begin{align}
g_{\alpha\beta}(\omega) ={}& g^>_{\alpha\beta}(\omega) + g^<_{\alpha\beta}(\omega)
\nonumber \\
={}& \sum_n\frac{\langle \psi_0^A| a_\alpha |\psi_n^{A+1}\rangle \langle\psi_n^{A+1}|a^\dagger_\beta|\psi_0^A\rangle}{\omega -(E_n^{A+1}-E_0^A)+i\eta}
\nonumber \\
& + \sum_f \frac{\langle \psi_0^A| a^\dagger_\beta |\psi_f^{A-1}\rangle \langle\psi_f^{A-1}|a_\alpha|\psi_0^A\rangle}{\omega -(E_0^{A}-E_f^{A-1})-i\eta}\, ,
\label{lehm:dec}
\end{align}
where $|\psi_0^A\rangle$ is the ground state wave function of $A$ nucleons, 
$|\psi_n^{A+1}\rangle$ ($|\psi_f^{A-1}\rangle$) are the eigenstates and $E_n^{A+1}$ ($E_f^{A-1}$)  the eigenvalues of the $(A\pm1)$-body system,
and $a^\dagger_\alpha$ and $a_\alpha$ are the creation and annihilation operator in the quantum state $\alpha$, respectively.

The one-body propagator given in Eq.~\eqref{lehm:dec} is completely determined by solving the Dyson equation
\begin{align}
g_{\alpha\beta}(\omega)=g^0_{\alpha\beta}(\omega)+\sum_{\gamma\delta}g^0_{\alpha\gamma}(\omega)\Sigma^\star_{\gamma\delta}(\omega)g_{\delta\beta}(\omega) \, ,
\label{dys:eq}
\end{align}
where $g^0_{\alpha\beta}(\omega)$ is the unperturbed single-particle propagator and $\Sigma^\star_{\gamma\delta}(\omega)$ is the irreducible self-energy that encodes nuclear medium effects in the particle propagator~\cite{Dickhoff2008book}. 
The latter is given by the sum of two different terms
\begin{align}
\Sigma^\star_{\alpha\beta}(\omega)=\Sigma^\infty_{\alpha\beta}+\tilde{\Sigma}_{\alpha\beta}(\omega)\ ,
\label{self:en}
\end{align}
the first one describes the average mean field while the second one contains dynamical correlations.
In practical calculations the self-energy is expanded as a function of the propagator itself, implying that
an iterative procedure is required to solve the Dyson equation self-consistently. 
Its dynamical part also has a Lehmann representation, which can be summarised as
\begin{align}
\tilde{\Sigma}_{\alpha\beta}(\omega)=\sum_{ij} {\bf D}_{\alpha i} \Big[\frac{1}{\omega-({\bf K}+{\bf C})} \Big]_{ij}{\bf D}^{\dagger}_{j \beta}\ ,
\label{self:en2}
\end{align}
where ${\bf K}$ are the unperturbed energies of 2p1h and 2h1p intermediate state configurations, ${\bf C}$ are interaction matrices among these configurations,
and ${\bf D}$ are coupling matrices to the single particle states.
We calculate Eq.\eqref{self:en} within the Algebraic Diagrammatic Construction (ADC) method, which consist in matching
the matrices $\Sigma^\infty_{\alpha\beta}$, ${\bf D}$ and ${\bf C}$ to the lowest terms in the perturbation theory expansion.
The third order truncation of this scheme [ADC(3)] yields a propagator that includes  all possible Feynman contributions up to third order
but it further resums infinite series of relevant diagrams in a non-perturbative fashion~\cite{Schirmer1983,Barbieri2017LNP}.
 The expressions of the static and dynamic self-energy up to third order, including all possible two- and three-nucleon terms that enter the expansion
 of the self-energy have been recently derived in Refs.~\cite{Carbone:2013eqa,Raimondi:2017kzi}.
In our calculations we use the intrinsic Hamiltonian (i.e., with the kinetic energy of the center of mass subtracted) including up to two- and three-nucleon forces (3NFs).
We reduce the number of Feynman diagrams that need to be considered by restricting the self-energy expansion to 
 only interaction-irreducible (\textit{i.e.} not averaged) diagrams~\cite{Carbone:2013eqa} and using (medium dependent) effective one- and two-body interactions.
The residual contributions due to pure  interaction-irreducible three-body forces is expected to be small and can be safely
neglected~\cite{Hagen:2007ew,Roth:2011vt,Barbieri2014QMBT,Cipollone:2014hfa}.

The poles and residues appearing in Eq.~\eqref{lehm:dec} provide the delta-function of energies and the nuclear transition amplitudes that enter the spectral function.
The nuclear matrix element entering Eq.~\eqref{pke:hole} are simply obtained transforming from the harmonic oscillator (HO) basis $\{\alpha\}$ (that we used for our calculations) to momentum space:
\begin{align}
[ \,\langle \psi_f^{A-1} | \otimes \langle k| ]  |  \psi_0^A\rangle &= \sum_\alpha {\cal Y}^k_\alpha \tilde{\Phi}_\alpha({\bf k}) \nonumber\\
&= \sum_\alpha \tilde{\Phi}_\alpha({\bf k}) \langle \psi_f^{A-1} | a_\alpha |\psi_0^A\rangle\, ,
\end{align}
and the more familiar expression of the spectral function written as the imaginary part of the hole Green's function becomes
\begin{align}
P_h(\mathbf{k},E)&=\frac{1}{\pi}\sum_{\alpha\beta}\tilde{\Phi}^\ast_\beta({\bf k})\tilde{\Phi}_\alpha({\bf k})
\;  \text{Im}  \left\{ g^<_{\alpha\beta}(\omega) \right\}  \, ,
%%%\nonumber\\ &\times\text{Im}\langle \psi_0^A| a_\beta^\dagger  \frac{1}{E+(H-E^A_0)-i\epsilon} a_\alpha|\psi_0^A\rangle\, ,
\label{eq:Ph}
\end{align}
where $\tilde\Phi_\alpha({\bf k})$ is the Fourier transform of the single-particle wave function
\begin{align}
\tilde{\Phi}_\alpha({\bf k})=\int d^3{r}\, e^{i {\bf k\, r}}\Phi_\alpha({\bf r})\, .
\label{HO:ft}
\end{align}

In this work, the SCGF calculations are performed employing a spherical HO basis, with frequency \hbox{$\hbar\Omega=20$~MeV} and dimension
\hbox{$N_{\rm max}={\rm max}\{2n+\ell\}=13$}. Within this basis we employ the NNLO$_{\rm sat}$ Hamiltonian, which was constructed following chiral perturbation
theory but fitted to reproduce radii and energies in mid mass nuclei~\cite{Ekstrom:2015rta}. Hence, it guarantees to reproduce the correct saturation point and
fundamental ground state properties of nuclei for masses in the region of $A\sim 12-40$.

The SCGF correlated one-body propagator obtained by solving the Dyson equation of Eq.~\eqref{dys:eq} is used to determine the hole SF of $^{16}$O in the ADC(3) approach. 
The results for open shell nuclei, such as $^{12}$C discussed in this work, have been obtained within the Gorkov's theory, in which the description of pairing 
correlations characterizing open shell systems is achieved by breaking the particle number symmetry~\cite{Soma:2011aj,Soma:2012zd,Soma:2013ona}.
However, Gorkov theory is currently only implemented up to second order [ADC(2)].

\subsection{Inclusion of two-body currents}
The inclusion of two-body current operator requires the generalization of the factorization {ansatz} of Eq.~\eqref{fact:1b}.
Following Refs.~\cite{Benhar:2015ula,Rocco:2015cil} and neglecting the contribution of $[\langle \psi_f^{A-1}|\otimes\langle p| ]|j^{\mu}_{2b} |\psi_0^A \rangle $,
the matrix element of the nuclear current reads  
\begin{align}
&\langle \psi_f^A|j^{\mu}_{2b} |\psi_0^A \rangle \to  \nonumber\\
&\qquad \sum_{k\,k^\prime} [\langle \psi_f^{A-2}| \otimes \langle k\, k^\prime | ] \, | \psi_0^A\rangle  _a\langle p\, p^\prime | \sum_{ij} j^\mu_{ij} | k\,k^\prime \rangle\, .
\label{fact:2b}
\end{align}
where $|p\,p^\prime\rangle_a=|p\,p^\prime\rangle-|p^\prime\,p\rangle$. In infinite matter the correlated nuclear many-body state can be labeled with their single-particle 
momenta, implying $|\psi_f^{A-2}\rangle = |h h^\prime\rangle$, where $|h h^\prime\rangle$ with $|\mathbf{h}|,|\mathbf{h}^\prime|\leq k_F$ denotes a 2-hole state of $(A-2)$ nucleons. 
A diagrammatic analysis of the cluster expansion of the overlap 
$\phi^{h h^\prime}_{k k^\prime}\equiv\langle \Psi_0 | [ |k k^\prime\rangle \otimes | \Psi_{h h^\prime}\rangle$ was carried out 
by the Authors of Ref.~\cite{Benhar:1999bg}. Their analysis shows that only unlinked graphs (i.e., those 
in which the points reached by the $k_1$, $k_2$ lines are not connected to one other by any dynamical or statistical correlation lines) 
survive in the $A \to \infty$ limit
\begin{equation}
\phi^{h h^\prime}_{k k^\prime} = \phi^{h }_{k} \phi^{h^\prime}_{k^\prime} (2\pi)^3\delta^{(3)}(\mathbf{h}-\mathbf{k})  (2\pi)^3\delta^{(3)}(\mathbf{h}^\prime-\mathbf{k}^\prime)\,,
\label{eq:2n_overlap}
\end{equation}
where $\phi^h_k$ is the the Fourier transform of the overlap between the ground state and the one-hole $(A-1)$-nucleon state,
the calculation of which is discussed in Ref.~\cite{Benhar:1989aw}

Therefore, using the $\delta^{(3)}$-function to perform the integration over ${\bf p}^\prime={\bf k}+{\bf k}^\prime+{\bf q}-{\bf p}$, the pure two-body 
current component of the hadron tensor in nuclear matter turns out to be~\cite{Benhar:2015ula}
\begin{align}
&W^{\mu\nu}_{\rm 2b}({\bf q},\omega)=\frac{V}{4} \int dE \frac{d^3k}{(2\pi)^3}  \frac{d^3k^\prime}{(2\pi)^3}\frac{d^3p}{(2\pi)^3}
\frac{m^4}{e({\bf k})e({\bf k^\prime})e({\bf p})e({\bf p^\prime})} \nonumber\\
 &\qquad \times  P_h^{\rm NM}({\bf k},{\bf k}^\prime,E) 2\sum_{ij}\, \langle k\, k^\prime | {j_{ij}^\mu}^\dagger |p\,p^\prime\rangle_a \langle p\,p^\prime |  j_{ij}^\nu | k\, k^\prime \rangle\nonumber\\
 &\qquad \times \delta(\omega+E-e(\mathbf{p})-e(\mathbf{p}^\prime))\, .
\end{align}

The normalization volume for the nuclear wave functions $V=\rho / A$ with $\rho=3\pi^2 k_F^3/2$ depends on the Fermi momentum of the nucleus,
which we take to be $k_F=225$ MeV. The factor $1/4$ accounts for the fact that we sum over indistinguishable pairs 
of particles, while the factor 2 stems from the equality of the product of the direct terms and the product of the two exchange terms after interchange
of indices~\cite{Dekker:1991ph}.
The two-nucleon SF entering the hadron tensor is 
\begin{align}
P_h^{\rm NM}(\mathbf{k},\mathbf{k}^\prime,E)&=\int \frac{d^3 h}{(2\pi)^3} \frac{d^3 h^\prime}{(2\pi)^3} |\phi^{h h^\prime}_{k k^\prime}|^2 \delta(E+e(\mathbf{h})+e(\mathbf{h}^\prime))\nonumber\\
&\times  \theta(k_F-|\mathbf{h}|) \theta(k_F-|\mathbf{h}^\prime|)\, .
\label{pke:hole2}
\end{align}
Consistently with the fact that, in absence of long-range correlations, the two-nucleon momentum distribution of infinite 
systems factorizes according to~\cite{Fantoni:1981jgs}
\begin{align}
n({\bf k},{\bf k}^\prime)=n({\bf k})n({\bf k}^\prime)+ {\cal O}\bigg(\frac{1}{A}\bigg)\, ,
\end{align}
exploiting the factorization of the two-nucleon overlaps of Eq.~\eqref{eq:2n_overlap}, the two-body contribution of the hadron tensor can be rewritten as 
\begin{align}
W^{\mu\nu}_{\rm 2b}({\bf q},\omega)&= \frac{V}{2}  \int d\tilde{E} \frac{d^3k}{(2\pi)^3} d\tilde{E}^\prime \frac{d^3k^\prime}{(2\pi)^3}\frac{d^3p}{(2\pi)^3}
 \nonumber\\
 &\times \frac{m^4}{e({\bf k})e({\bf k^\prime})e({\bf p})e({\bf p^\prime})} P_h^{\rm NM}({\bf k},\tilde{E})P_h^{\rm NM}({\bf k}^\prime,\tilde{E}^\prime)\nonumber\\
&\times  \sum_{ij}\, \langle k\, k^\prime | {j_{ij}^\mu}^\dagger |p\,p^\prime \rangle \langle p\,p^\prime |  j_{ij}^\nu | k\, k^\prime \rangle\nonumber\\
& \times \delta(\omega+\tilde{E}+\tilde{E}^\prime-e(\mathbf{p})-e(\mathbf{p}^\prime))\, .
\label{had:tens2}
\end{align}
In order to make contact with finite systems, we take 
\begin{equation}
P_h^{\rm NM}({\bf k},E) \simeq  \frac{k_F^3}{6\pi^2} P_h({\bf k},E)
\end{equation}
where the hole SF of the nucleus $P_h({\bf k},E)$ is obtained from either the CBF theory or the SCGF approach.  

We are aware that the assumptions made to include the contribution of two-body currents deserve further investigations. For instance, the strong isospin-dependence 
of short-range correlations, elucidated in a number of recent works~\cite{Wiringa:2013ala,Weiss:2015mba,Weiss:2018tbu}, is not properly accounted for if the
factorization of Eq.~\eqref{eq:2n_overlap}. In this regard, it has 
to be mentioned that in the present work we do not account for the interference between one- and two-body currents. While in the two-nucleon knockout 
final states this contribution is relatively small~\cite{Benhar:2015ula,Rocco:2015cil}, it has been argued that tensor correlations strongly enhance 
the interference terms for final states  associated single-nucleon knock out processes~\cite{Fabrocini:1996bu}. This is consistent with the Green's function 
Monte Carlo calculations of Refs.~\cite{Lovato:2013,Lovato:2014}, in which the interference between one- and two-body currents dominate the total
two-body current contribution.

\section{Electroweak current operators}
\label{sec:currents}
We analyze the neutrino- and anti-neutrino- nucleus 
quasielastic scattering induced by both CC and NC transitions. 
The elementary interactions for the CC processes are
\begin{align}
\nu(k)+n(p) &\to \ell^-(k^\prime) + p(p^\prime)\ ,\\
{\bar \nu}(k)+p(p)&\to \ell^+(k^\prime) + n(p^\prime)\,,
\end{align}
while for NC transitions 
\begin{align}
\label{NC:p}
\nu(k)+p(p)&\to \nu(k^\prime) + p(p^\prime)\, ,\\   
\label{NC:n}
\nu(k)+n(p)&\to \nu(k^\prime) + n(p^\prime)\, .
\end{align}
The corresponding ones for the anti-neutrino are obtained replacing 
$\nu$ with $\bar{\nu}$ both in the initial and final states. 

The one-body current operator is the sum of vector (V) and axial (A) terms
 for both CC and NC processes and it can be written as
\begin{align}
&j^\mu= (J^\mu_V+J^\mu_A) \nonumber\\
&J^\mu_V={\cal F}_1 \gamma^\mu+ i \sigma^{\mu\nu}q_\nu \frac{{\cal F}_2}{2M}\nonumber\\
&J^\mu_A=-\gamma^\mu \gamma_5 {\cal F}_A -q^\mu \gamma_5 \frac{{\cal F}_P}{M}\ .
\end{align}
The Conserved Vector Current (CVC) hypothesis allows to relate
the vector form factor to the electromagnetic ones. 
For CC processes they are given by
\begin{align}
{\cal F}_i &=F^p_i-F^n_i\, ,
\end{align}
where 
\begin{align}
F_1^{p,n}&=\frac{G_E^{p,n}+\tau G_M^{p,n}}{1+\tau}\nonumber\\
F_2^{p,n}&=\frac{G_M^{p,n}-G_E^{p,n}}{1+\tau}
\end{align}
with $\tau=-q^2/4M^2$. As for the proton and neutron electric and magnetic form factors,
we adopted the Galster parametrization~\cite{Galster:1971kv}
\begin{align}
G_E^p&=\frac{1}{(1-q^2/M_V^2)^2}\ ,\ &G_M^p=\mu_p G_E^p\nonumber \\
G_E^n&= -\frac{\mu_n \tau}{(1+\lambda_n\tau)}G_E^p\ ,\ &G_M^n=\mu_n G_E^p
\end{align}
with $M_V=0.843$ GeV, $\mu_p= 2.7928$, $\mu_n=-1.9113$, and $\lambda_n=-5.6$.
In this work we neglect the pseudoscalar form factor ${\cal F}_P=F_P$, since in the cross
section formula it is multiplied by the mass of the outgoing lepton.
As for the axial form factor ${\cal F}_A=F_A$, we assume the standard dipole parametrization
\begin{align}
 F_A &=\frac{g_A}{( 1- q^2/ M_A^2 )^2}\ ,
\end{align}
where the nucleon axial-vector coupling constant is taken to be $g_A=1.2694$~\cite{PDG} and the axial mass 
$M_A=1.049$ GeV. The dipole parametrization of $F_A$ has been the subject of
intense debate and an alternative ``z-expansion'' analyses~\cite{Meyer:2016oeg} has recently been proposed. Understanding
how the $q^2$ dependence of the axial form factor impact predictions for the neutrino cross sections,
in particular relatively to uncertainties in modeling nuclear dynamics, is certainly interesting, and will be 
investigated in future works.

The single-nucleon form factors relevant to the NC neutrino-proton scattering of Eq.~\eqref{NC:p},
read
\begin{align}
{\cal F}_i=&\Big(\frac{1}{2}-2 \sin^2\theta_W\Big)F^p_i-\frac{1}{2}F^n_i-\frac{1}{2}F^s_i\, ,\nonumber\\
{\cal F}_A=&\frac{1}{2}F_A+\frac{1}{2}F_A^s\, ,
\end{align}
while those relevant for the NC neutrino-neutron scattering process of Eq.~\eqref{NC:n} are
\begin{align}
{\cal F}_i=&\Big(\frac{1}{2}-2 \sin^2\theta_W\Big)F^n_i-\frac{1}{2}F^p_i-\frac{1}{2}F^s_i\, ,\nonumber\\
{\cal F}_A=&-\frac{1}{2}F_A+\frac{1}{2}F_A^s\, ,
\end{align}
where $\theta_W$ is the Weinberg angle ($\sin^2\theta_W = 0.2312$~\cite{PDG}). The form factors
$F^s_i$ and $F_A^s$ describe the strangeness content of the nucleon. Following Ref.~\cite{Leitner:2008ue}, 
we set
\begin{align}
F^s_i&=0\, ,\nonumber\\
F_A^s&=-\frac{0.15}{(1-q^2/M_A^2 )^2}\, .
\end{align}

\begin{figure}[]
\centering
\includegraphics[scale=0.5]{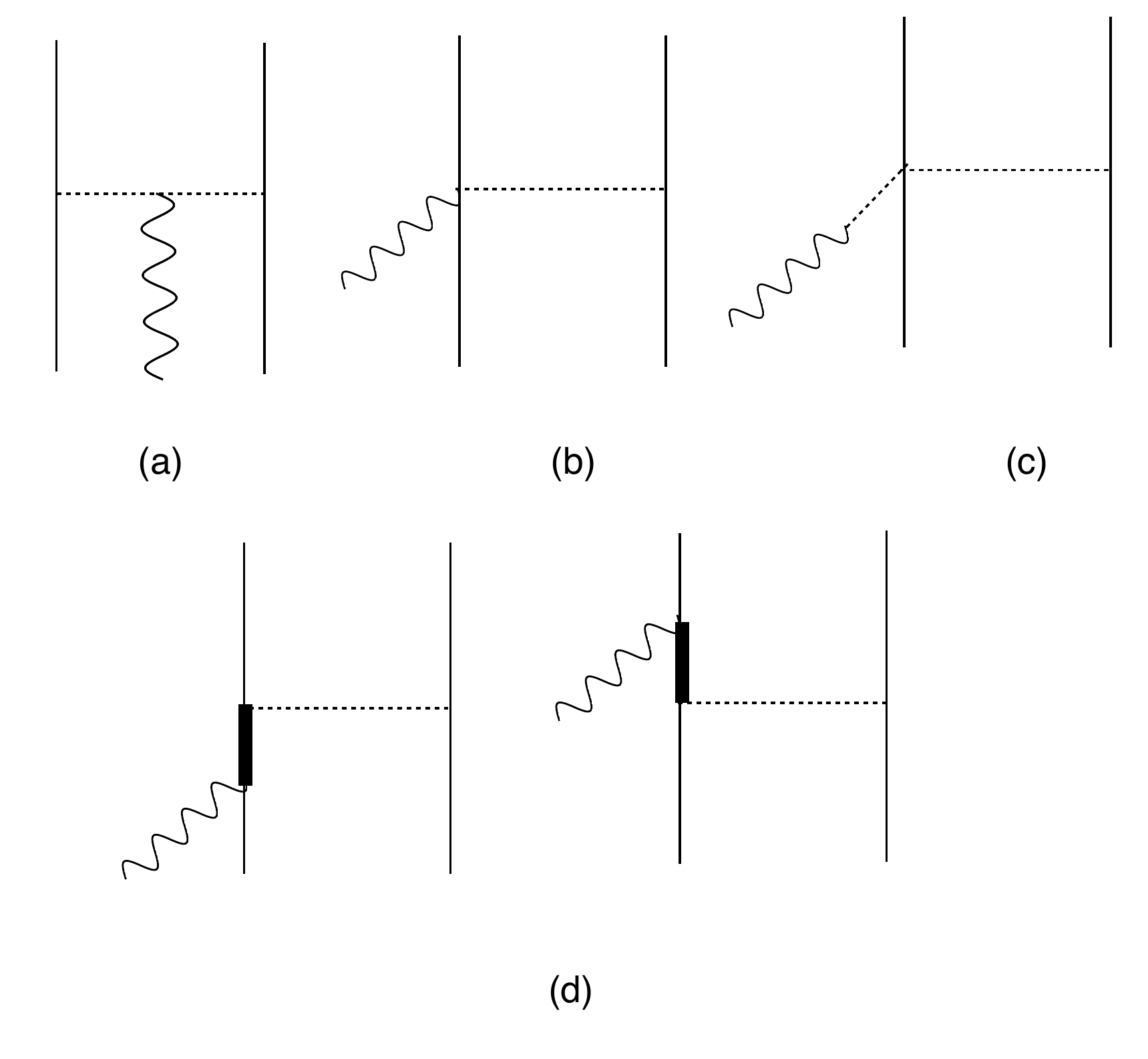}
\caption{Ferynman diagrams describing two-body currents contributions associated to: pion in flight (a), seagull (b), pion-pole (c),
and delta excitations (d) processes. Solid, thick, and dashed lines correspond to nucleons, deltas, pions, respectively. The wavy
line represents the vector boson.}
\label{mec:diag}
\end{figure}

The electroweak meson exchange current operators used in our work are those employed in Ref.~\cite{Simo:2016ikv}. 
They have been derived by coupling the pion-production amplitudes obtained within the
non-linear $\sigma$ model in Ref.~\cite{Hernandez:2007qq} to a second nucleon line. 
The meson exchange current operator is the sum of four different contributions
\begin{align}
j^\mu_{\rm MEC}= j^\mu_{\pi}+j^\mu_{\rm sea}+ j^\mu_{\rm pole}+ j^\mu_{\Delta}\, ,
\end{align}
whose corresponding Feynman diagrams are depicted in Fig.~\ref{mec:diag}.

Introducing the pion momenta $k_1=p-k$ and $k_2=p^\prime - k^\prime$, the 
pion-in-flight current operator corresponding to diagram (a) of Fig.~\ref{mec:diag}
is written as
\begin{align}
j^\mu_\pi & =(I_V)_\pm J^\mu_\pi \ ,\nonumber\\
J^\mu_\pi &=(J^\mu_\pi)_V+(J^\mu_\pi)_A\ ,\nonumber\\
(J^\mu_\pi)_V &= \frac{f^2_{\pi NN}}{m_\pi^2}F_1^V(q) F_{\pi NN}(k_1)F_{\pi NN}(k_2) \nonumber\\
&\times \Pi(k_1)_{(1)}\Pi(k_2)_{(2)}(k_1^\mu-k_2^\mu)\  ,\nonumber\\
(J^\mu_\pi)_A&= 0\ .
\label{pif:curr}
\end{align}
where $f^2_{\pi NN}/(4\pi)$=0.08 and the pion propagation and absorption is described by
\begin{align}
\Pi(k)=\frac{\gamma_5\slashed{k}}{k^2-m_\pi^2}\, .
\end{align}
The isospin raising-lowering operator is given by
\begin{align}
(I_V)_\pm=(\tau^{(1)}\times \tau^{(2)})_\pm\ ,
\end{align} 
where $\pm\rightarrow x\pm iy$.

To preserve the CVC, in the vector part of the pion-in-flight current operator we include
the electromagnetic form factor
\begin{align} 
F_1^V(q)=G_E^p(q)-G_E^n(q)\, .
\label{fpi:vec}
\end{align}
The $\pi NN$ coupling is described using a form factor that accounts for the off-shellness of the
pion
\begin{align}
&F_{\pi NN}(k)= \frac{\Lambda_\pi^2-m_\pi^2}{\Lambda_\pi^2-k^2}\label{fpinn}\ ,
\end{align}
where $\Lambda_\pi$=1300 MeV.

The electroweak seagull current operator, given by the sum of diagram (b) of Fig. \ref{mec:diag} and
the one obtained interchanging particles 1 and 2, reads
\begin{align}
j^\mu_{\rm sea}&=(I_V)_\pm J^\mu_{\rm sea} \ ,\nonumber\\
J^\mu_{\rm sea}&=(J^\mu_{\rm sea})_V+(J^\mu_{\rm sea})_A\ ,\nonumber\\
(J^\mu_{\rm sea})_V&= \frac{f^2_{\pi NN}}{m_\pi^2}F_1^V(q) F^2_{\pi NN}(k_1) \Pi(k_1)_{(1)} \big (\gamma_5 \gamma^\mu\big)_{(2)} \nonumber\\
&- (1\leftrightarrow 2)\ ,\nonumber\\
(J^\mu_{\rm sea})_A&=\frac{f^2_{\pi NN}}{m_\pi^2}\frac{1}{g_A}F_\rho(k_2) F^2_{\pi NN}(k_1) \Pi(k_1)_{(1)} (\gamma^\mu\big)_{(2)} \nonumber\\
& - (1\leftrightarrow 2)\, .
\label{sea:curr}
\end{align}
The form factor $F_\rho(k)$, included to account for the $\rho$ meson dominance of the $\pi NN$ coupling,
is given by 
\cite{Hernandez:2007qq}
\begin{align}
F_\rho(k)=\frac{1}{k^2-m^2_\rho}, \ \ \  m_\rho=775.8\ {\rm MeV}
\end{align}

The expression for the pion-pole current operator, represented by diagram (c) of Fig. \ref{mec:diag}, is 
\begin{align}
j^\mu_{\rm pole}&=(I_V)_\pm J^\mu_{\rm pole} \ ,\\
J^\mu_{\rm pole}&=(J^\mu_{\rm pole})_V+(J^\mu_{\rm pole})_A\ , \\
(J^\mu_{\rm pole})_V&=0 ,\\
(J^\mu_{\rm pole})_A&= \frac{f^2_{\pi NN}}{m_\pi^2}\frac{1}{g_A}F_\rho(k_1)F^2_{\pi NN}(k_2) \Pi(k_2)_{(2)} \nonumber\\
&\times  \Big(\frac{q^\mu\slashed{q}}{q^2-m_\pi^2}\Big)_{(1)}- (1\leftrightarrow 2)\ .
\label{pole:curr}
\end{align}
Diagrams (d), as well as the corresponding two in which particles 1 and 2 are interchanged, are associated with
 two-body current terms involving a $\Delta$-resonance in the intermediate state.
The expression of this operator is largely model dependent, owing to the purely transverse nature of this current, {\textit i.e.}
the form of the vector part  is not subject to current-conservation constraints. We adopted the parametrization of Ref.~\cite{Hernandez:2007qq}
\begin{align}
j^\mu_{\Delta}&=\frac{3}{2}\frac{f_{\pi NN} f^\ast}{m^2_\pi} \bigg\{ \Pi(k_2)_{(2)}
\Big[ \Big( -\frac{2}{3}\tau^{(2)}+\frac{I_V}{3}\Big)_{\pm} \nonumber\\
&\times F_{\pi NN}(k_2) F_{\pi N \Delta} (k_2) (J^\mu_a)_{(1)}-\Big(\frac{2}{3}\tau^{(2)}+\frac{I_V}{3}\Big)_{\pm} \nonumber\\
&\times F_{\pi NN}(k_2) F_{\pi N \Delta} (k_2) (J^\mu_b)_{(1)}\Big]+(1\leftrightarrow 2)\bigg\}
\label{delta:curr}
\end{align}
where $f^\ast$=2.14 and 
\begin{equation}
F_{\pi N \Delta}(k)=\frac{\Lambda^2_{\pi N\Delta}}{\Lambda^2_{\pi N\Delta}-k^2}\ ,
\end{equation}
with $\Lambda_{\pi N\Delta}=1150$ MeV.
The $N\rightarrow \Delta$ transition vertices entering the left and right (d) diagrams, corresponding to $J^\mu_a$ and $J^\mu_b$, respectively
are expressed as
\begin{align}
J^\mu_a&=(J^\mu_a)_V+(J^\mu_a)_A\ ,\nonumber\\
(J^\mu_a)_V&=\frac{C_3^V}{M}\Big[k_2^\alpha G_{\alpha\beta}(h_1+q)\Big(g^{\beta\mu}\slashed{q}-q^\beta\gamma^\mu\Big)\Big]\gamma_5\ ,\nonumber\\
(J^\mu_a)_A&=C_5^A \Big[k_2^\alpha G_{\alpha\beta}(h_1+q) g^{\beta\mu}\Big]
\end{align}
and
\begin{align}
J^\mu_b&=(J^\mu_b)_V+(J^\mu_b)_A\ ,\nonumber\\
(J^\mu_b)_V&=\frac{C_3^V}{M}\gamma_5\Big[\Big(g^{\alpha\mu}\slashed{q}-q^\alpha\gamma^\mu\Big)G_{\alpha\beta}(p_1-q)k_2^\beta\Big],\nonumber\\
(J^\mu_b)_A&=C_5^A \Big[ g^{\alpha\mu} G_{\alpha\beta}(p_1-q) k_2^\beta\Big]\, .
\end{align}
Since the above $\Delta$ current is applied in the resonance region, the standard Rarita-Schwinger propagator
\begin{align}
G^{\alpha\beta}(p_\Delta)=\frac{P^{\alpha\beta}(p_\Delta)}{p^2_\Delta-M_\Delta^2}
%G^{\alpha\beta}(p_\Delta)=\frac{P^{\alpha\beta}(p_\Delta)}{p^2_\Delta-M_\Delta^2+i M_\Delta \Gamma_\Delta(p_\Delta)}
\end{align}
has to be modified to account for the possible $\Delta$  decay into a physical $\pi N$ state. To this aim, following Refs.~\cite{Dekker:1994yc,DePace:2003}, 
we replaced the real resonance mass $M_\Delta$=1232 MeV by $M_\Delta - i \Gamma(p_\Delta)/2$. The energy-dependent 
decay width $\Gamma(p_\Delta)/2$ effectively accounts for the allowed phase space for the pion produced in the physical decay process. It is
given by
\begin{equation}
\Gamma(p_\Delta)=\frac{(4 f_{\pi N \Delta})^2}{12\pi m_\pi^2} \frac{|\mathbf{k}|^3}{\sqrt{s}} (m_N + E_k) R(\mathbf{r}^2)
\end{equation}
where $(4 f_{\pi N \Delta})^2/(4\pi)=0.38$, $s=p_\Delta^2$ is the invariant mass, $\mathbf{k}$ is the decay three-momentum
in the $\pi N$ center of mass frame, such that
\begin{equation}
\mathbf{k}^2=\frac{1}{4s}[s-(m_N+m_\pi)^2][s-(m_N-m_\pi)^2]\,
\end{equation} 
and $E_k=\sqrt{m_N^2 + \mathbf{k}^2}$ is the associated energy. The additional factor
\begin{equation}
R(\mathbf{r}^2)=\left(\frac{\Lambda_R^2}{\Lambda_R^2-\mathbf{r}^2}\right)
\end{equation}
depending on the $\pi N$ three-momentum $\mathbf{r}$, with $\mathbf{r}^2=(E_k - \sqrt{m_\pi^2 + \mathbf{k}^2})^2-4\mathbf{k}^2$ and $\Lambda_R^2=0.95\, m_N^2$,
is needed to better reproduce the experimental phase-shift $\delta_{33}$~\cite{Dekker:1994yc}.
In addition, to avoid double-counting with real pion emission, as in Refs.~\cite{DePace:2003,Simo:2016ikv,Butkevich:2017mnc} we only keep the real part of the $\Delta$ propagator.
The spin 3/2 projection operator reads
\begin{align}
P^{\alpha\beta}(p_\Delta)&=(\slashed{p}_\Delta+M_\Delta)\Big[ g^\alpha\beta -\frac{1}{3}\gamma^\alpha\gamma^\beta  -\frac{2}{3}\frac{p_\Delta^\alpha p_\Delta^\beta}{M_\Delta^2}\nonumber\\
&+\frac{1}{3}\frac{p_\Delta^\alpha \gamma^\beta - p_\Delta^\beta \gamma^\alpha}{M_\Delta}\Big]\ .
\end{align}
%The energy dependent decay width reads
%\begin{align}
%\Gamma_\Delta(s)&=\frac{1}{6\pi}\Big(\frac{}{}

The vector and axial form factors adopted in this work are those of Ref.~\cite{Hernandez:2007qq}
\begin{align}
C_3^V&=\frac{2.13}{(1-q^2/M_V^2)^2}\ \frac{1}{1-q^2/(4 M_V^2)}\ ,\label{cv3:del}\\
C_5^A&=\frac{1.2}{(1-q^2/M^2_{A\Delta})^2}\ \frac{1}{1-q^2/(3M^2_{A\Delta})}\ ,\label{c5a:del}
\end{align}
where $M_V=0.84$ GeV and $M_{A\Delta}=1.05$ GeV.

The MEC employed here are purely isovector. Hence, the currents relevant
to NC processes are obtained by replacing the $\pm \rightarrow z$ component in the isospin operator, for example
\begin{equation}
(I_V)_{\pm}\rightarrow (I_V)_z=(\tau^{(1)}\times\tau^{(2)})_z\ .
\end{equation}
Following the discussion of Ref.~\cite{Leitner:2008ue} we rewrite the vector form factors of Eqs.~\eqref{fpi:vec},~\eqref{cv3:del} as
\begin{align}
\tilde{F}_V&=(1-2\sin\theta_W^2) F_V\ ,\\
\tilde{C}_3^V&=(1-2\sin\theta_W^2)C_3^V\ ,
\end{align}
while the axial form factors are the same as in the CC case.
%%%%%%%%%%%%%%%%%%%%%%%%%%%%%%%%
\section{Numerical implementation}
\label{sec:numerical}

\begin{figure}[]
\centering
\includegraphics[width=\columnwidth]{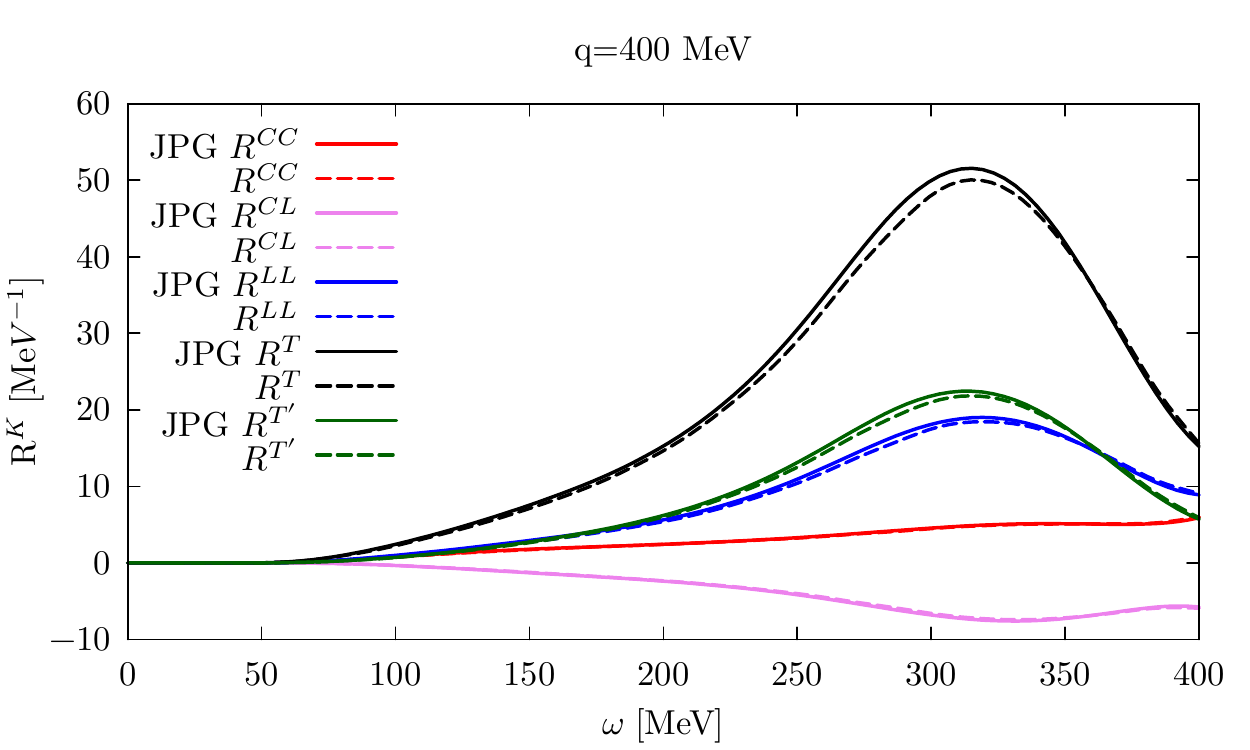}
\includegraphics[width=\columnwidth]{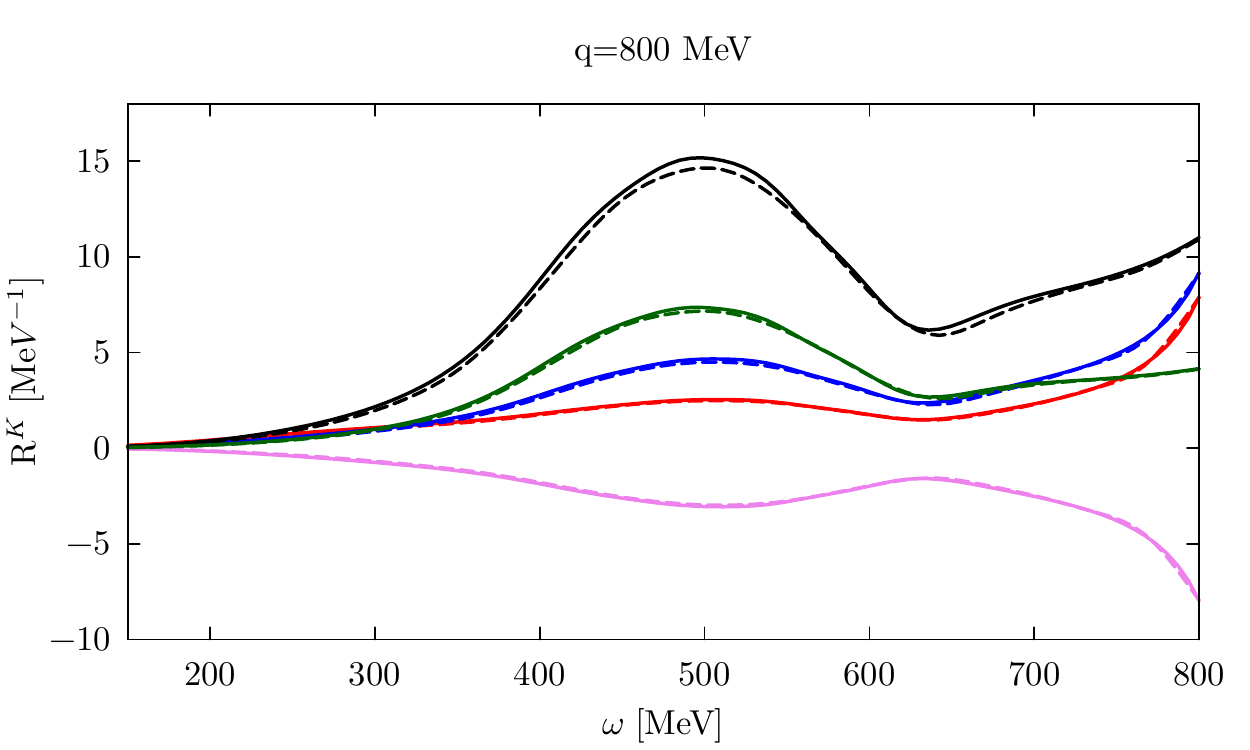}
\caption{ The upper and lower panel display the two-body CC response functions of $^{12}$C for $|{\bf q}|$=400, and 800 MeV,
 respectively, obtained within the Relativistic Fermi gas model. We benchmark our results displayed by the dashed curves with those of Ref.~\cite{Simo:2016ikv}  corresponding to the solid curves.  }
\label{res:amaro:12b}
\end{figure}

The large number of terms entering the current operator defined in Eqs.~\eqref{pif:curr}, \eqref{sea:curr}, \eqref{pole:curr}, and \eqref{delta:curr} greatly complicates the calculation of the 
two-body response functions. Explicitly summing the matrix elements of the two-body currents over the initial and final spin states gives rise to thousands of terms, the inclusion of which involves non-trivial difficulties. To overcome them, we developed Fortran subroutines able to automatically compute the required matrix elements performing an explicit spin-isospin summation. We note that this procedure allows for a straightforward inclusion of the exchange terms, avoiding the complications encountered by the Authors of Refs.~\cite{Dekker:1994yc,DePace:2003,Rocco:2015cil}. 

Taking  $\mathbf{p}=(p\cos \theta_p ,0,p\sin \theta_p )$ as in Ref.~\cite{Dekker:1994yc} and using the energy-conserving delta function to determine $p$, the eleven-dimensional integral of Eq.~\eqref{had:tens2} can 
be reduced to a nine-dimensional one, schematically written as
\begin{align}
W^{\mu\nu}_{\rm 2b}({\bf q},\omega)&= \int dX I^{\mu\nu} (X, {\bf q},\omega)\, .
\label{eq:int_9d}
\end{align}
In the above equation we have introduced the generalized coordinate $X=\{\mathbf{k},\mathbf{k}^\prime,\tilde{E},\tilde{E}',\cos \theta_p\}$, while the integrand is given by
\begin{align}
& I^{\mu\nu}(X, {\bf q},\omega)=\frac{m^4}{e({\bf k})e({\bf k^\prime})e({\bf p})e({\bf p^\prime})} P_h^{\rm NM}({\bf k},\tilde{E}) P_h^{\rm NM}({\bf k}^\prime,\tilde{E}^\prime)\nonumber\\
&\qquad \times \frac{p^2}{(2\pi)^8}  \sum_{ij}\, \langle k\, k^\prime | {j_{ij}^\mu}^\dagger |p\,p^\prime \rangle \langle p\,p^\prime |  j_{ij}^\nu | k\, k^\prime \rangle\,
\end{align}
It has long been known that Monte Carlo methods provide an efficient way to evaluate large-dimensional integrals. In this regard, let us express the integral of Eq.~\eqref{eq:int_9d} as
\begin{equation}
W^{\mu\nu}_{\rm 2b}({\bf q},\omega)=  \int dX \mathcal{P} (X) \frac{I^{\mu\nu}(X)}{\mathcal{P} (X)}
\end{equation}
where $\mathcal{P} (X)$ is a probability distribution. According to the central limit theorem, the above integral can be estimated by sampling a sequence of points $X_{i}$ distributed according to $\mathcal{P} (X)$
\begin{align}
W^{\mu\nu}_{\rm 2b}({\bf q},\omega) & \simeq \frac{1}{N_X} \sum_{X_i} \frac{I^{\mu\nu}(X_i)}{\mathcal{P} (X_i)}\, .
\end{align}
with $N_X$ being the number of Monte Carlo samples. Its variance decreases asymptotically to zero as $1/N_X$, regardless the number of integration variables
\begin{align}
\sigma_{W^{\mu\nu}_{\rm 2b}}^2 ({\bf q},\omega) &\simeq  \frac{1}{N_X(N_X-1)} \left[  \sum_{X_i}  \left( \frac{I^{\mu\nu}(X_i)}{\mathcal{P} (X_i)}\right)^2 \right. \nonumber\\
& - \left. \left(\sum_{X_i} \frac{I^{\mu\nu}(X_i)}{\mathcal{P} (X_i)}\right)^2 \right]\, .
\end{align}

\begin{figure*}[!]
\includegraphics[scale=0.5]{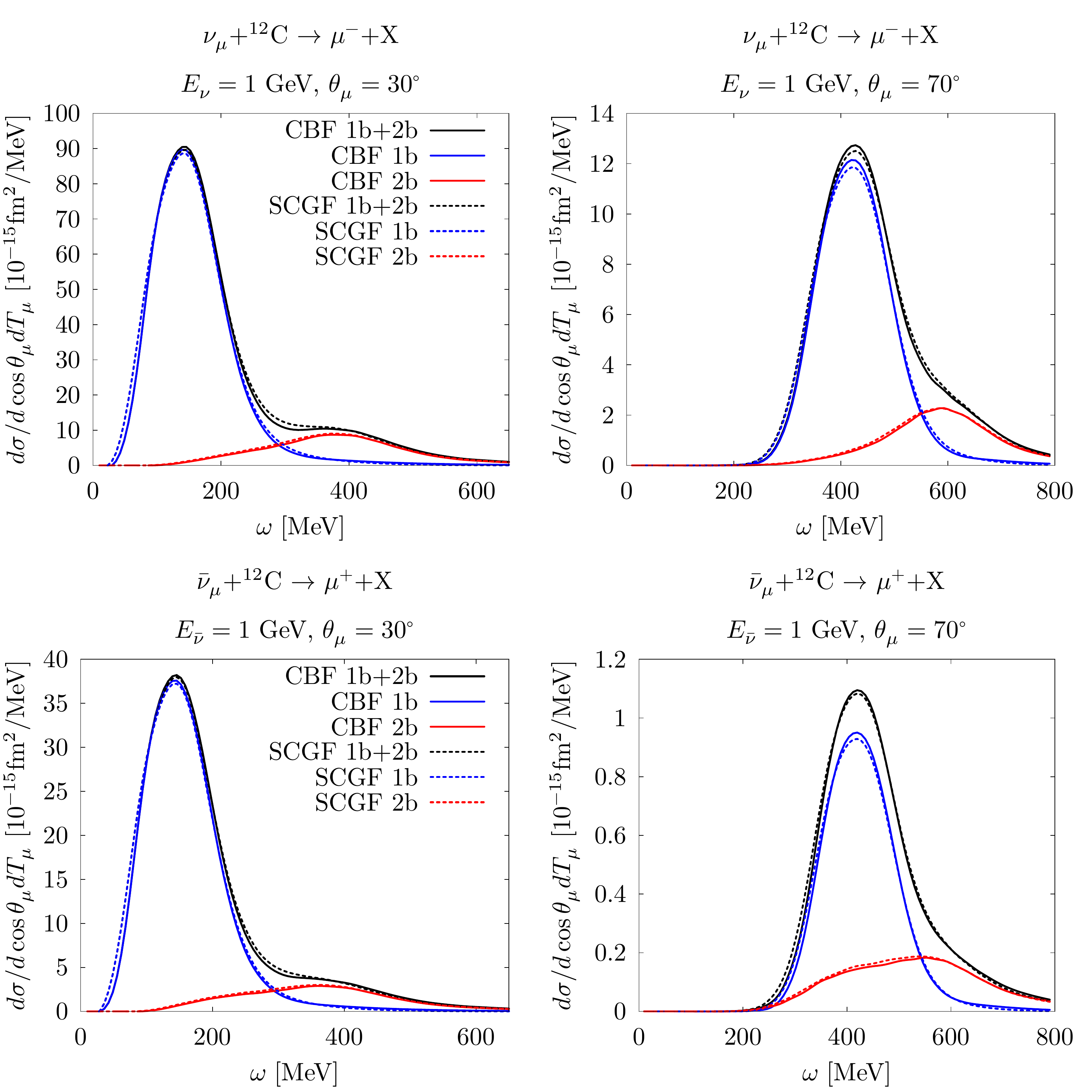}
\caption{ The upper panels correspond to the CC inclusive differential cross section of $\nu_\mu$ scattering
on $^{12}$C for $E_\nu$=1 GeV and $\theta_\mu=30^\circ$ and $70^\circ$, respectively. 
The blue (red) lines correspond to including only one-body  (two-body) contributions in the CC reaction,
while the black lines displays the total result.
Dotted lines show results from the SF computed with the SCGF method and solid lines are from CBF.
The lowest panels are the same as the upper but for $\bar{\nu}_\mu$ scattering on $^{12}$C.}
\label{cross:sec:CC:C12}
\end{figure*} 

The SFs employed in this work include the contribution of correlated pairs of nucleon, hence they extend up to large momentum and removal energy. As a consequence, the phase space spanned by the nucleons in the initial state is significantly larger than in the Fermi-gas case. To efficiently perform the integral of Eq.~\eqref{eq:int_9d}, we chose the following normalized importance-sampling function
\begin{equation}
\mathcal{P}(X)=\frac{1}{2} \frac{k_F^6}{(6\pi)^2} P_h^{\rm NM}({\bf k},\tilde{E}) P_h^{\rm NM}({\bf k}^\prime,\tilde{E}^\prime)\, .
\end{equation}
We generate the sequence of points $X_i$ sampling $\mathcal{P}(X)$ according to the Metropolis algorithm~\cite{Metropolis:1953am}. Exploiting the importance-sampling allows to achieve a percent-level precision with $N_X\sim 5 \times 10^6$. Note that, to reduce autocorrelation of samples, we compute the integral every $10$ steps, so that the total number of samples in the Monte Carlo path is $5 \times 10^7$. We take full advantage of the fact that Monte Carlo algorithms are known to be ``embarrassingly parallel''. Our calculations are distributed over dozens of MPI ranks reaching an almost ideal efficiency, as very little communication between the different ranks is required. More specifically, computing the five response functions relevant for neutrino-nucleus scattering for a given value of momentum transfer requires less than one minute of computing time on 64 Intel Xeon E5-2600 (Broadwell) processors. 

Our integration algorithm presents a number of advantages with respect to the standard deterministic methods usually employed in the calculation of the nuclear response function. For instance, we neither employ the so-called ``frozen approximation''  -- amounting to neglect the momenta of the two initial nucleons~\cite{Simo:2014wka} -- nor we need to parametrize the response functions before computing the double-differential and total cross sections~\cite{Megias:2014qva,Megias:2016fjk}.

Considering the SF of a uniform isospin-symmetric Fermi gas of nucleons with Fermi momentum $k_F=225$ MeV, we benchmarked our results for the two-body charged-current responses of $^{12}$C against those of Ref.~\cite{Simo:2016ikv}, obtained within the relativistic Fermi gas model using the same current operators. The remarkably good agreement between the two  calculations, displayed in Fig.~\ref{res:amaro:12b} for $|{\bf q}|=400$ MeV and $|{\bf q}|=800$ MeV, considerably corroborates their accuracy. It has to be stressed that achieving this degree of consistency for such elaborate calculations must not be taken for granted. In fact, the models of Refs.~\cite{Martini:2011wp,Nieves:2011yp}, although based on similar models of nuclear dynamics, differ in about a factor of two in their estimation of the size of the multi-nucleon effects~\cite{Nieves:2017kqt}.

\begin{figure*}[!]
\includegraphics[scale=0.5]{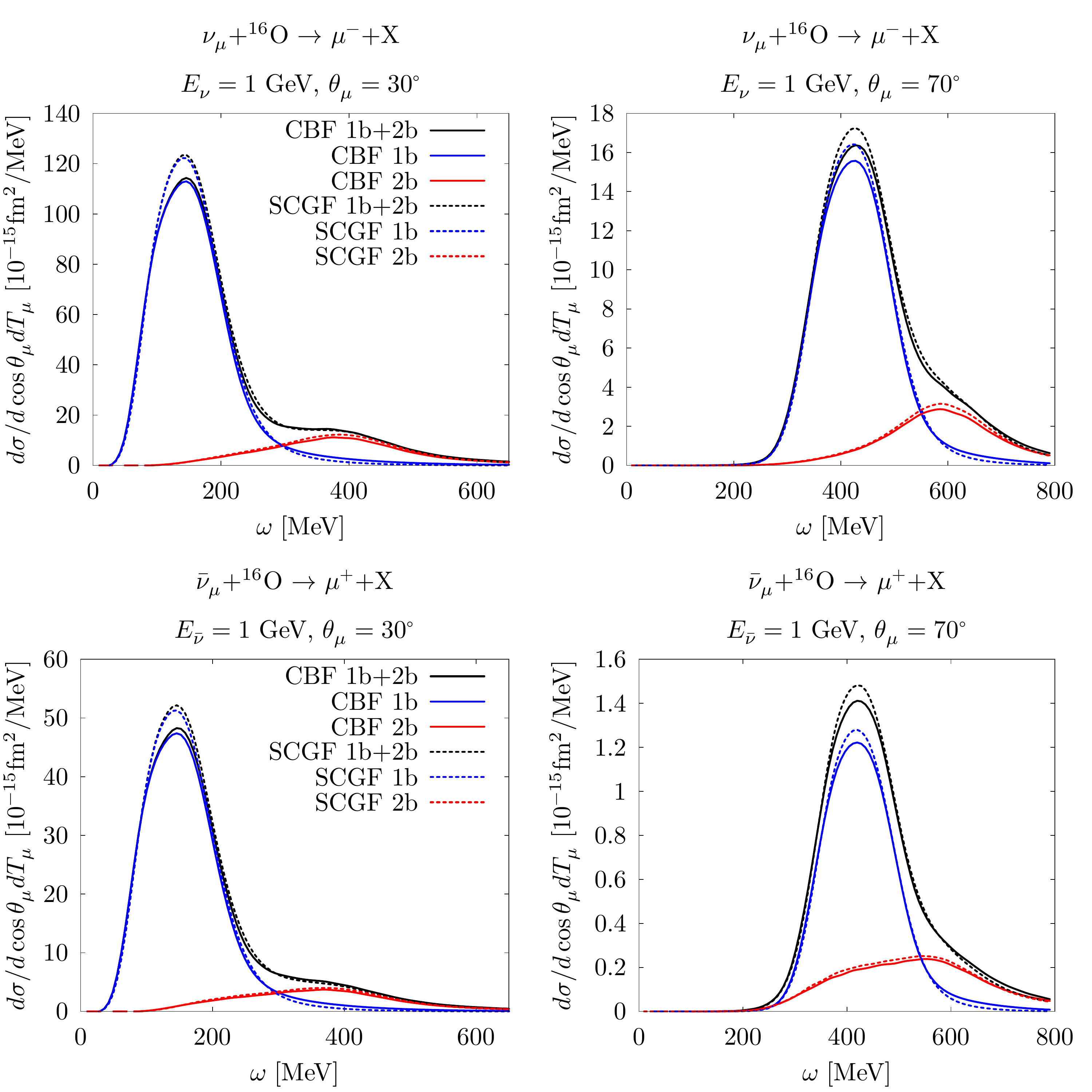}
\caption{ The upper panel correspond to the CC inclusive differential cross section of $\nu_\mu$ scattering
on $^{16}$O for $E_\nu$=1 GeV and $\theta_\mu=30^\circ$ and $70^\circ$, respectively. 
The blue (red) lines correspond to including only one-body  (two-body) contributions in the CC reaction,
while the black lines displays the total result.
Dotted lines show results from the SF computed with the SCGF method and solid lines are from CBF.
The lowest panels are the same as the upper but for $\bar{\nu}_\mu$ scattering
on $^{16}$O.}
\label{cross:sec:CC:O16}
\end{figure*} 

Analogously to the electromagnetic case, two-body currents are most effective in the transverse channels. On the other hand, we observe a non-negligible enhancement in $R^{CC}$ and $R^{LL}$, driven by the axial two-body pieces of the current operator, consistently with the findings of Refs.~\cite{Kubodera:1978wr,Lovato:2017cux}.

\section{Results}
\label{sec:results}
In this Section we present our findings for neutrino and anti-neutrino scattering off $^{12}$C and $^{16}$O nuclei, for both CC and NC reactions,
gauging the differences between the hole SFs discussed in Sec.~\ref{sec:impulse}. It has to be noted that the CBF SF relies 
on the semi-phenomenological AV18+UIX Hamiltonian, which naturally encompass short-range correlations. On the other hand, the softer 
NNLO$_{\rm sat}$ interaction is adopted in the SCGF approach. Hence, our analysis might serve as a comparison
between two distinctive models of nuclear dynamics.
In this preliminary study, we neglect FSI between the struck nucleon(s) and the 
spectator systems. They will be accounted for when flux-folded doubly-differential cross sections will be computed, which will require a separate
publication. 

\begin{figure*}[!]
\includegraphics[scale=0.5]{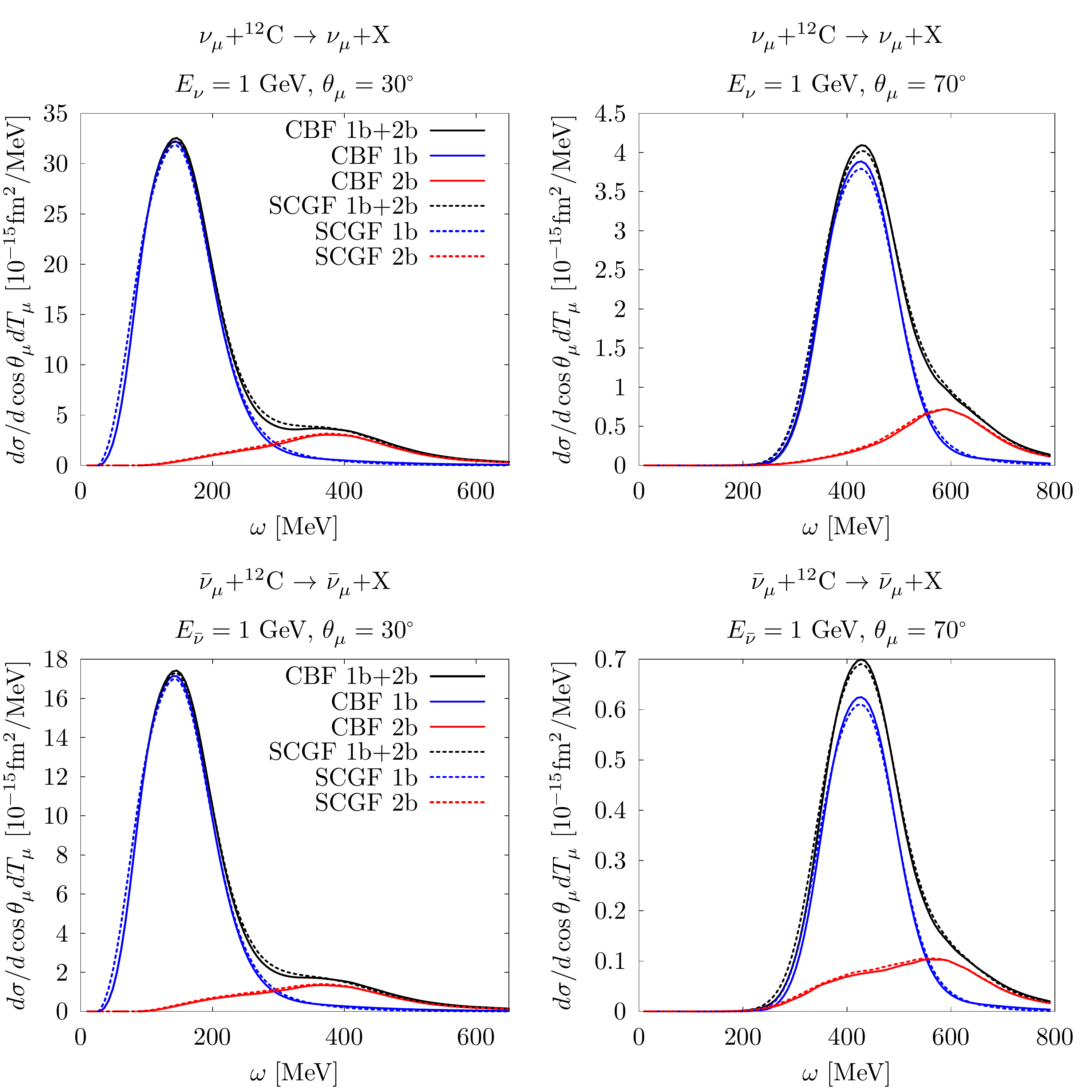}
\caption{Same as for Fig.~\ref{cross:sec:CC:C12} but for the NC inclusive differential cross sections. }
\label{cross:sec:NC:C12}
\end{figure*} 

The upper panels of Fig.~\ref{cross:sec:CC:C12} show the $\nu_\mu$-$^{12}$C inclusive differential cross section 
induced by CC transitions for $E_\nu$=1 GeV and $\theta_\mu=30^\circ$ (left panel) and $\theta_\mu=70^\circ$ (right panel).
The solid and the dashed  curves have been obtained employing the CBF and SCGF hole SFs, respectively. The full calculations, which include
both one- and two-body currents, are displayed by the black lines. The red and blue curves separately highlight one- and two-body current 
contributions. The lower panel is analogous to the upper one but for $\bar{\nu}_\mu$-$^{12}$C scattering processes. 

Calculations carried out employing the CBF and SCGF hole SFs are in remarkably good agreement, although they are obtained from different,
albeit realistic, input Hamiltonians.
Consistently with the findings of Ref.~\cite{Rocco:2015cil}, two-body currents primarily enhance the cross-sections in the ``dip region", between the 
quasielastic peak and the resonance-production region. The excess strength provided by meson-exchange currents increases relatively
to the total cross section for larger values of the scattering angle. In fact, as shown in Fig.~\ref{res:amaro:12b}, two-body contributions are 
most effective in the transverse responses, although this feature is less clearcut than in the electromagnetic case. It has to be noted that, 
in the anti-neutrino case, for $\theta_\mu=70^\circ,$ two-body currents are also effective for quasielastic kinematics.
Because of the cancellation in Eq.~\eqref{eq:cross_sec} between the contributions proportional to the $R_{T}$ and $R_{T^\prime}$ responses, 
the anti-neutrino cross section decreases rapidly relatively to the neutrino cross section as the scattering angle changes from $\theta_\mu=30^\circ$ 
to $\theta_\mu=70^\circ$.

Figure~\ref{cross:sec:CC:O16} is analogous to Fig.~\ref{cross:sec:CC:C12} but for $\nu_\mu$- and $\bar{\nu}_\mu$-$^{16}$O scattering. 
For this closed-shell isotope, the self-energy can be computed in the ADC(3) truncation of the SCGF approach, which includes
all-order resummations of phonons.
Hence, the propagator is more accurate than that of an open-shell nucleus as $^{12}$C. In addition, since $^{16}$O comprises more nucleons than $^{12}$C, 
the LDA entering the CBF calculation of the hole SF is expected to be more reliable. Nonetheless, a comparison between the solid and dashed curves reveals somewhat larger
discrepancies between the CBF and SCGF results than in the $^{12}$C case.
As shown in Figs.~\ref{cross:sec:CC:C12} and~\ref{cross:sec:CC:O16}, the SCGF one-body cross-sections exhibit an enhancement in the
peak region and a (feeble) quenching of the high-energy transfer tails with respect to the corresponding CBF predictions. This is consistent with the fact 
that the chiral nuclear potential employed in the SCGF approach is softer than the one included in the CBF formalism, as highlighted in the analysis of the 
single-nucleon momentum distributions carried out in Ref.~\cite{Rocco:2018vbf}. 
It has long been known that short-range correlations in accurate semi-phenomenologic
potentials lead to a quenching of the mean-field strength of the SF by about~10\%~\cite{Muether1995prcO16src,VanNeck1998prcCMO16,Dickhoff2004ppnp}.
Although the SCGF spectroscopic factors computed from NNLO$_{\rm sat}$ describe rather well the quenching observed in $(p,2p)$ knockout, they are slightly 
higher than the empirical $(e,e'p)$ ones encoded in our CBF calculations. 

One may interpret the discrepancies between the CBF and SCGF results as a (crude) indication of the theoretical uncertainty. However, a rigorous estimate
of the latter requires to employ electroweak currents that are consistent with the two models of the nuclear Hamiltonian, as well as a more accurate treatment 
of FSI.

\begin{figure*}[]
\includegraphics[scale=0.5]{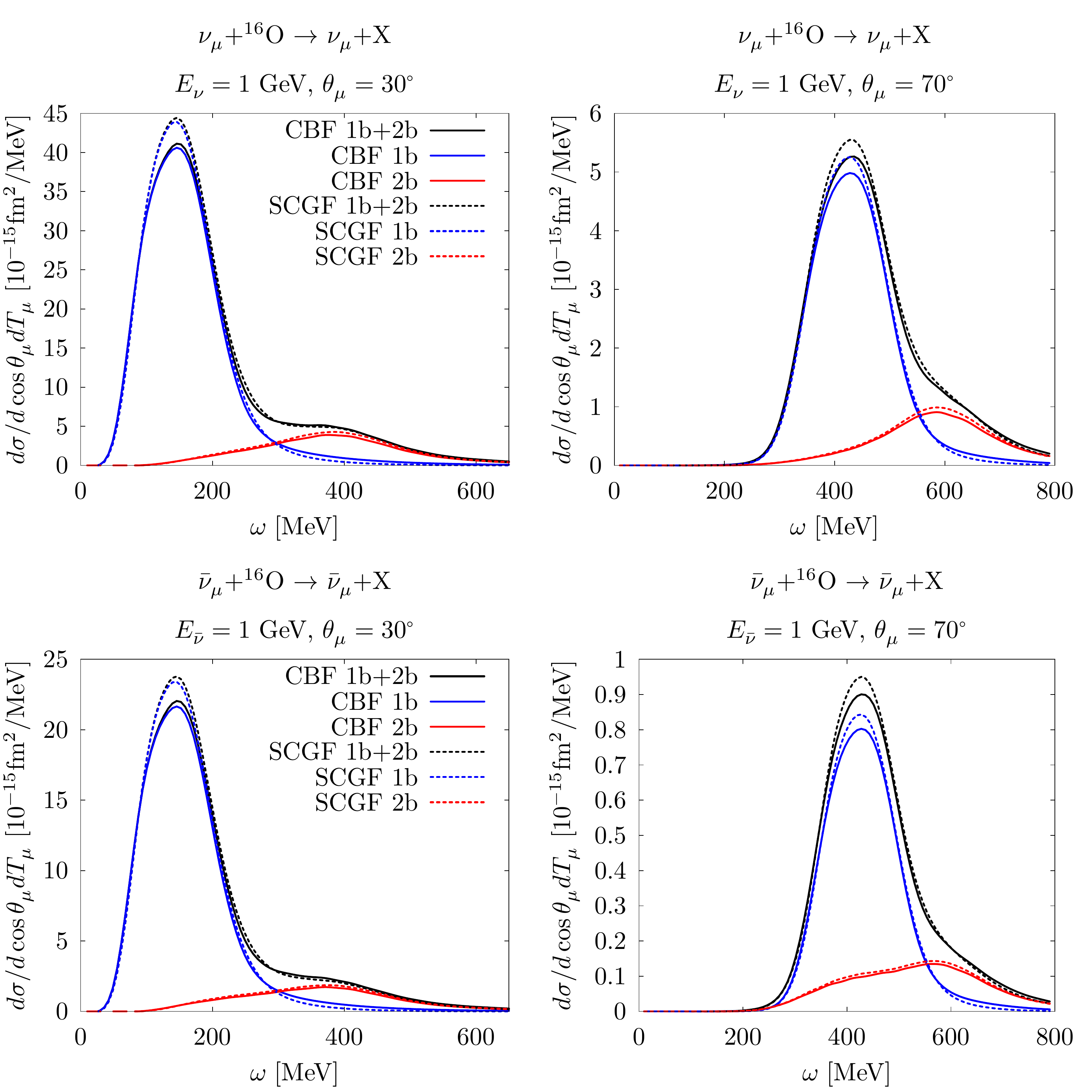}
\caption{Same as for Fig.~\ref{cross:sec:CC:O16} but for the NC inclusive differential cross sections. }
\label{cross:sec:NC:O16}
\end{figure*}

The results for the NC double differential cross sections of ${\nu}_\mu$ and $\bar{\nu}_\mu$ scattering off
$^{12}$C and $^{16}$O nuclei are displayed in Fig.~\ref{cross:sec:NC:C12} and Fig.~\ref{cross:sec:NC:O16}, respectively.
We consider the same kinematics as before, namely $E_{\nu}=1$ GeV and  $\theta_\mu= 30^\circ$ and $\theta_\mu= 70^\circ$. 
There is an overall good agreement between the CBF and SCGF predictions, particularly apparent for the
$^{12}$C nucleus, as already observed for CC transitions. Consistently with the CC case, two-body terms mostly affect the dip
region, although for anti-neutrino scattering and $70^\circ$ they also provide excess strength in the quasielastic-peak region.

\begin{figure}[]
\centering
\includegraphics[width=\columnwidth]{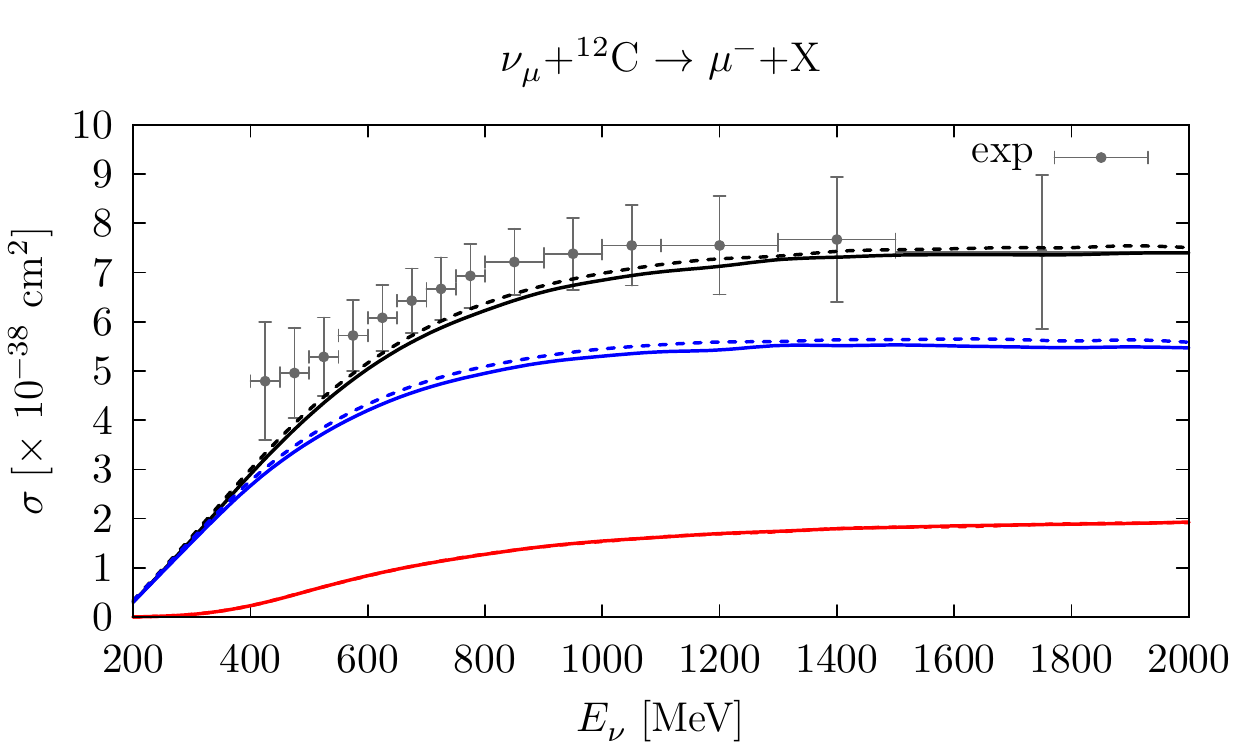}
\includegraphics[width=\columnwidth]{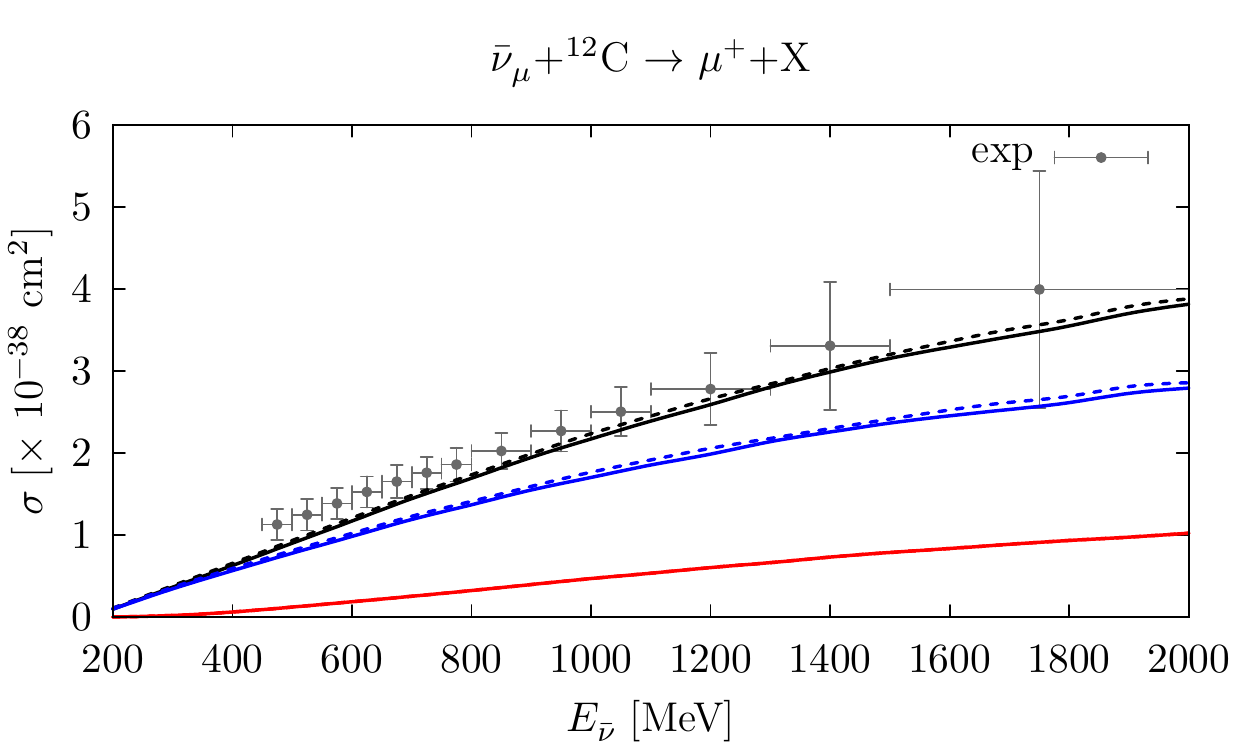}

\caption{CCQE $\nu_\mu$-$^{12}$C total cross section per nucleon as a function of the neutrino energy.
The blue (red) lines correspond to including only one-body (two-body) contributions in the CC reaction, while the black lines displays the total result. Dotted lines show results from the SF computed with the SCGF method and solid lines are from CBF.
 The MiniBooNE data points ~\cite{AguilarArevalo:2010zc} are plotted as a function of the reconstructed neutrino energy. }
\label{total:cross:sec:CC:C12}
\end{figure}

In Fig.~\ref{total:cross:sec:CC:C12} we display the total cross section per nucleon as a function of the neutrino energy,
compared to the values extracted from the analysis carried out by the MiniBooNE collaboration~\cite{Aguilar:2008,AguilarArevalo:2010zc}. Consistently with our findings
relative to the double-differential cross sections, MEC substantially increase one-body results over the entire range of incoming neutrino
energy. We also note that the curves referring to the CBF and SCGF hole SFs are almost superimposed, a further validation of the
robustness of our predictions. The overall good agreement with experimental values, achieved once that two-body currents are accounted
for, must not be overrated, for at least two main reasons. When reconstructing the incoming energy,
a relativistic Fermi gas is employed in the event-generator and only one-body scattering processes are accounted for. It has been 
argued that both two-body currents~\cite{Ankowski:2014yfa,Ankowski:2016bji,Nieves:2012yz} and a realistic description of the target state are likely to alter the reconstructed value of 
$E_{\nu,\bar{\nu}}$. In addition, the MiniBooNE analysis of the data corrects (through a Monte Carlo estimate) for some of these events, where in
the neutrino interaction a real pion is produced, but it escapes detection because it is reabsorbed in the nucleus, leading to multi-nucleon emission.

\section{Conclusions}
\label{sec:conclusions}
In this work we have set the stage to include relativistic MEC currents relevant for both CC and NC transitions within realistic models of nuclear
dynamics. We studied their behavior in neutrino and anti-neutrino scattering off $^{12}$C and $^{16}$O nuclei, which constitute the targets 
of current~\cite{Aguilar:2008,Gran:2006jn,Lyubushkin:2008pe}, and next-generation~\cite{dune_web} neutrino-oscillation experiments. In this regard, we computed the double-differential cross sections 
for incoming energy of $E_{\nu,\bar{\nu}}=1$ GeV and two values of the scattering angle: $\theta_\mu= 30^\circ$ and $\theta_\mu= 70^\circ$. The total cross section 
for neutrino and anti-neutrino $^{12}$C scattering has also been evaluated and compared with the values extracted by the MiniBooNE collaboration.

We use the relativistic meson-exchange currents originally derived in Ref.~\cite{Hernandez:2007qq} to describe pion-production processes. Subsequently, 
these currents were implemented in the relativistic Fermi gas model to account for two-particle two-hole final state channels in 
electron- and neutrino-nucleus scattering~\cite{Simo:2016ikv}. Calculations performed combining this contribution to the SUSAv2 prediction for the
quasielastic region show that the inclusion of MEC appreciably improves the agreement with electron- and neutrino-nucleus scattering data~\cite{Megias:2014qva,Megias:2016fjk,Megias:2016lke}. 

We developed an highly-optimized parallel code, based on the Metropolis Monte Carlo algorithm, to efficiently evaluate the NC and CC cross 
sections and response functions. As for the latter, within the Fermi gas model we have carried out a successful comparison with the results 
reported in Ref.~\cite{Simo:2016ikv} for two values of the momentum transfer that supports the correctness of both calculations. Capitalizing on medium-size 
computer clusters allows us to avoid approximations, such as the {\it frozen nucleon} one, often adopted when employing deterministic integration 
procedures~\cite{RuizSimo:2017onb,RuizSimo:2017hlc}. In addition, when computing neutrino-nucleus cross sections, we do not make use of {\it ad hoc} parameterizations of the 
response functions~\cite{Megias:2014qva,Megias:2016lke}. 

In order to combine a realistic description of the target nucleus with relativistic currents and kinematics, we employ the formalism based on factorization
using realistic hole SFs,  and  follow the scheme devised in Ref.~\cite{Benhar:2015ula,Rocco:2015cil} to account for two-nucleon emission processes. The required 
nuclear amplitudes and the consistent hole SFs are obtained from two different many body schemes, and using different models of nuclear dynamics.

The CBF theory
and the SCGF approach, both rely upon a non-relativistic nuclear Hamiltonian  to describe the interactions among nucleons. 
However, the phenomenological Hamiltonian employed to perform the CBF calculation has been derived from a fit of the properties of
the {\it exactly solvable} two- and three-nucleon systems\textemdash including the measured scattering phase-shifts at laboratory energies
up to 300 MeV\textemdash and fails to provide an accurate description of the spectra and radii of nuclei with $A>4$~\cite{Lonardoni:2017egu}.
The chiral Hamiltonian employed in the SCGF calculation, on the other hand, is designed to reproduce the the properties of light and
medium-mass nuclei~\cite{Ekstrom:2015rta}, but fails to describe nucleon-nucleon scattering above 35 MeV. 
It has to be pointed out that the procedure followed to obtain the NNLO$_{\rm sat}$ potential implies a significant departure from the so-called
{\it ab initio} approach, in which the determination of nuclear dynamics is decoupled from the theoretical uncertainty associated with the calculation
of nuclear observables for  $A>4$.
In spite of these limitations, predictions of radii, charge form factors and spectral quantities from NNLO$_{\rm sat}$ 
are found to be in very good agreement with the experimental data~\cite{Ruiz2015Nature,Lapoux:2016exf,Lecluse2017Si34,Atar2018prl}, 
corroborating the use of this interaction to investigate the electroweak response functions of medium-mass isotopes.

In view of the above observations, the interpretation of the substantial agreement between the CC and NC cross-sections obtained from the two approaches, 
without adjusting any parameters, is not straightforward. 
It is interesting to note that, despite the inability of the NNLO$_{\rm sat}$ to reproduce the phase shifts at high energies, the SCGF SF predicts a high energy 
tail of the cross section, reflecting the presence in the wave function of momentum components in the range $200-400$~MeV.
This is clearly visible in Fig. 8 of Ref.~\cite{Rocco:2018vbf},  where the SCGF single-nucleon momentum-distributions are shown to be compatible, up to 
relatively large momenta, with those obtained using  Quantum Monte Carlo techniques and the AV18+UIX potential.

Consistently with Refs.~\cite{Benhar:2015ula,Rocco:2015cil}, we found that, for CC transitions, MEC provide excess strength primarily in the dip region.
Only for the larger value of the scattering angles we considered, $\theta_\mu= 70^\circ$, and for anti-neutrino processes, we find that two-body
currents enhance the quasielastic peak region. A similar behavior is also observed for NC-induced processes, somehow at variance with
the GFMC results of Ref.~\cite{Lovato:2017cux}. There, MEC were found to significantly increase the NC cross section for quasielastic kinematics, primarily
because of the interference between the one-and two-body current matrix element. The latter process, which was found to be relatively small
for two-nucleon knockout final states, has been disregarded in the present analysis. The interference term and FSI will be accounted for in the
forthcoming calculation of the flux-folded double-differential cross section, which allows for a more direct comparison with experimental data.

This work represents a significant step forward towards the realization of the strategy, advocated by the Authors of Ref.~\cite{Benhar:2010} to
describe neutrino-nucleus scattering in the whole kinematical region relevant for neutrino-oscillation experiments. In this regards, it has to be
noted that the same formalism used in this work is suitable to consistently describe the resonance-production region; work 
in this direction, aimed at extending the results of Ref.~\cite{Vagnoni:2017hll}, is underway. Both the CBF and the SCGF approaches are currently being 
developed to tackle the formidable problem of neutrino scattering off 
$^{40}$Ar; preliminary electron-scattering results obtained with the SCGF hole SF are encouraging. 

\section{Acknowledgements}
Conversations and e-mail exchanges with J.E. Amaro and I. Ruiz Simo are gratefully acknowledged. This research is supported by the U.S. 
Department of Energy, Office of Science, Office of Nuclear Physics, under contracts DE-AC02-06CH11357 (A.L. and N.R.),
by Fermi Research Alliance, LLC, under Contract No. DE-AC02-07CH11359 with the U.S. Department of Energy, Office of Science, Office of High Energy Physics (N. R.),
by the Centro Nazionale delle Ricerche (CNR) and the Royal Society under the CNR-Royal Society International Fellowship Scheme No. NF161046 (N. R.),
by the United Kingdom Science and Technology Facilities Council (STFC) under Grants No. ST/P005314/1 and No. ST/L005816/1 (C. B.).
Numerical calculations have been made possible through a CINECA-INFN  agreement, providing access to resources on MARCONI at CINECA. 
The computation of SCGF SFs was performed using the DiRAC Data Intensive service at Leicester (funded by the UK BEIS via STFC capital
grants ST/K000373/1 and ST/R002363/1 and STFC DiRAC Operations grant ST/R001014/1).

\bibliography{biblio}
\end{document}